\documentclass[conference]{IEEEtran}
\IEEEoverridecommandlockouts
\usepackage{cite}
\usepackage{amsmath,amssymb,amsfonts}
\usepackage{graphicx}
\usepackage{textcomp}
\usepackage{xcolor}
\usepackage{physics}

\usepackage{subfigure}
\usepackage{algorithm}
\usepackage[noend]{algpseudocode}
\usepackage{graphicx}
\usepackage{textcomp}
\usepackage{xcolor}
\usepackage{booktabs}
\usepackage{hyperref} 
\usepackage{url}
\usepackage{amsmath}
\usepackage{algpseudocode}
\usepackage{multirow}

\def\BibTeX{{\rm B\kern-.05em{\sc i\kern-.025em b}\kern-.08em
    T\kern-.1667em\lower.7ex\hbox{E}\kern-.125emX}}
\begin{document}

% \title{Trigger generation technique for Black-Box targeted attack on sensor based HAR}
\title{Dynamic Black-box Backdoor Attacks on IoT Sensory Data}
% {\footnotesize \textsuperscript{*}Note: Sub-titles are not captured in Xplore and
% should not be used}
% \thanks{Identify applicable funding agency here. If none, delete this.}
% }

% \author{\IEEEauthorblockN{Ajesh Koyatan Chathoth\IEEEauthorrefmark{1}\IEEEauthorrefmark{2} and
% Stephen Lee\IEEEauthorrefmark{1}}
% \IEEEauthorblockA{\IEEEauthorrefmark{1}University of Pittsburgh,
% \IEEEauthorrefmark{2}Eaton Corporation}}

\author{\IEEEauthorblockN{Ajesh Koyatan Chathoth and Stephen Lee}
\IEEEauthorblockA{
% \textit{Department of Computer Science} \\
\textit{University of Pittsburgh, Pittsburgh, USA}
% \\
% Pittsburgh, USA
}
% \and
% \IEEEauthorblockN{Stephen Lee}
% \IEEEauthorblockA{
% % \textit{Department of Computer Science} \\
% \textit{University of Pittsburgh}\\
% Pittsburgh, USA }
}

% %\and
% %\IEEEauthorblockN{Stephen Lee}
% % \IEEEauthorblockA{
% % % \textit{Department of Computer Science} \\
% % \textit{University of Pittsburgh}\\
% % Pittsburgh, USA }
% }
% \and
% \IEEEauthorblockN{3\textsuperscript{rd} Given Name Surname}
% \IEEEauthorblockA{\textit{dept. name of organization (of Aff.)} \\
% \textit{name of organization (of Aff.)}\\
% City, Country \\
% email address or ORCID}
% \and
% \IEEEauthorblockN{4\textsuperscript{th} Given Name Surname}
% \IEEEauthorblockA{\textit{dept. name of organization (of Aff.)} \\
% \textit{name of organization (of Aff.)}\\
% City, Country \\
% email address or ORCID}
% \and
% \IEEEauthorblockN{5\textsuperscript{th} Given Name Surname}
% \IEEEauthorblockA{\textit{dept. name of organization (of Aff.)} \\
% \textit{name of organization (of Aff.)}\\
% City, Country \\
% email address or ORCID}
% \and
% \IEEEauthorblockN{6\textsuperscript{th} Given Name Surname}
% \IEEEauthorblockA{\textit{dept. name of organization (of Aff.)} \\
% \textit{name of organization (of Aff.)}\\
% City, Country \\
% email address or ORCID}
% }

\maketitle

\begin{abstract}
Sensor data-based recognition systems are widely used in various applications, such as gait-based authentication and human activity recognition (HAR).
Modern wearable and smart devices feature various built-in Inertial Measurement Unit (IMU) sensors, and such sensor-based measurements can be fed to a machine learning-based model to train and classify human activities.  
While deep learning-based models have proven successful in classifying human activity and gestures, they pose various security risks.
In our paper, we discuss a novel dynamic trigger-generation technique for performing black-box adversarial attacks on sensor data-based IoT systems. Our empirical analysis shows that the attack is successful on various datasets and classifier models with minimal perturbation on the input data. We also provide a detailed comparative analysis of performance and stealthiness to various other poisoning techniques found in backdoor attacks.
We also discuss some adversarial defense mechanisms and their impact on the effectiveness of our trigger-generation technique.
 %, fall detection, physical training, sports coaching, gaming, driver activity, alerting systems, and security systems. 
\end{abstract}


\section{Introduction}
Smart devices, equipped with advanced sensors and connectivity, are enabling new and emerging applications in mobile sensing. From tracking physical activity to monitoring health conditions via gait analysis, these devices transform our interaction with our environments. For instance, by leveraging sensor data, these devices can recognize users or even diagnose health conditions~\cite{de2018sensor, pantelopoulos2009survey}. Meanwhile,  recent advances in deep learning have significantly enhanced the accuracy and utility of smart device applications, driving increased research interest in their potential uses~\cite{xu2019innohar, thu2021hihar}. As deep learning and sensor technologies continue to evolve, we expect more widespread use of deep learning in diverse smart device applications.

Despite the popularity of deep learning, there is a growing security concern regarding its application~\cite{goodfellow2014explaining, gu2019badnets}. Deep neural network (DNN) models are particularly vulnerable to backdoor attacks, where attackers design specific triggers that cause the model to misclassify inputs containing those triggers. For example, an attacker could insert a distinctive accessory, such as a unique pair of glasses, as a trigger, which could deceive a face recognition model into granting access to anyone wearing similar glasses~\cite{sharif2019general}. In fact, recent advancements have made these triggers more stealthy, making them harder to detect~\cite{zhou2024stealthy}. 

While there has been considerable progress in understanding and mitigating backdoor attacks in various domains, such as vision and natural language processing, sensor-based applications have received comparatively less attention~\cite{wang2019neural}. Smart devices often rely on time-series data from sensors like accelerometers and gyroscopes, which capture dynamic patterns of human movement over time. This time-dependent nature introduces additional complexities, such as the need to detect subtle changes in activity patterns and manage noise in sensor data. Consequently, sophisticated backdoor attacks that exploit these temporal dynamics remain a critical but underexplored area~\cite{ding2022towards}. Such attacks could involve subtle manipulations of sensor data sequences, which are challenging to detect but could lead to significant vulnerabilities in smart device-based systems. Addressing these research gaps is essential for enhancing the security of various smart Internet of Things (IoT) applications.

In this paper, we address the issue of backdoor attacks on deep learning models used in IoT sensor data, focusing on generating backdoor triggers dynamically to ensure their stealthiness. Our key research question is how to introduce triggers in sensor data dynamically so that the attack remains covert and difficult to detect.  Additionally, we examine how practical DNN backdoor attacks can be conducted in a black-box setting, where the model’s internal workings are inaccessible.  Prior studies in domains such as computer vision have demonstrated that fixed triggers --- triggers that do not change based on input --- are relatively easy to detect~\cite{liu2017neural}. This is because these fixed triggers exhibit consistent patterns that can be identified by defense mechanisms through pattern recognition techniques. For example, a fixed trigger might involve a specific watermark or pattern that is present in the training data and can be detected during the model's operation~\cite{li2022backdoor}. In contrast, dynamic triggers --- those that change from one input to another --- present a much greater challenge for detection. These triggers adapt based on the input data, which makes them harder for traditional detection methods to recognize. As a result, dynamic triggers can significantly enhance the stealthiness of backdoor attacks, making them more difficult to identify and defend against.

\begin{figure}[ht]
  \centering
  \includegraphics[width=2.8in]%{figures/Generatoronly_pdelta.png}
  {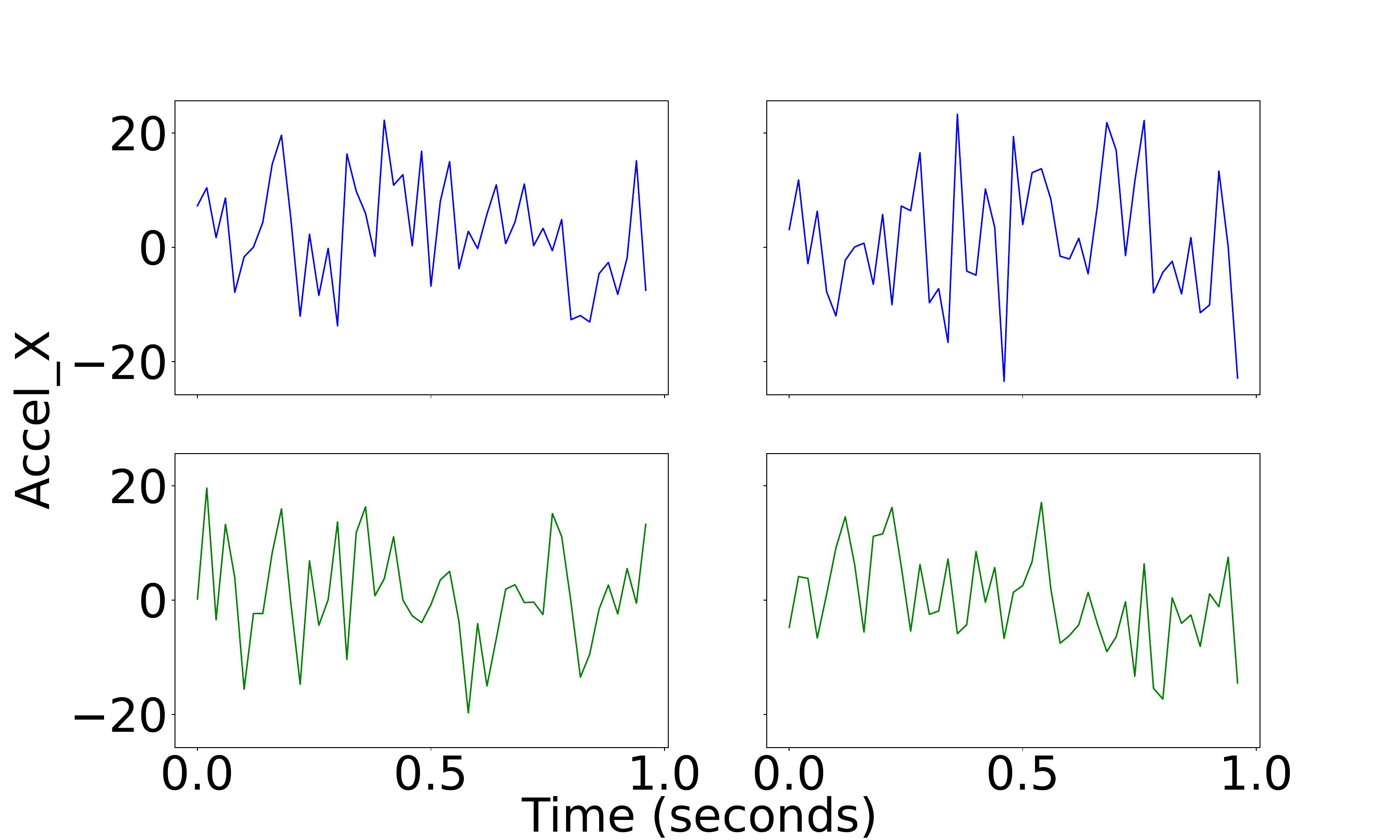}
  \caption{Accelerometer samples of Gait dataset of two participants at two different time periods. Top row: Person A, Bottom row: Person B.}
  \vspace{-3mm}
  %\Description{Perturbation - Delta.}
  \label{fig:samplefig}
\end{figure}

Generating dynamic triggers for time-series sensor data in a black-box setting is particularly challenging due to several inherent factors. Time-series data involves temporal dependencies that evolve over time, making it significantly harder to apply fixed patterns compared to static images or text. Unlike static data, time-series data requires triggers that can continuously adapt to remain effective. Additionally, sensor data exhibits considerable variability across different tasks and individuals (see Figure~\ref{fig:samplefig}) and can be influenced by environmental factors, necessitating a robust approach to handle these variations. 
This challenge is further exacerbated in a black-box setting, where the deep neural network (DNN) model's weights are inaccessible and unmodifiable. In this scenario, attackers must identify effective triggers that exploit vulnerabilities in the model using only observable input-output interactions. Without direct access to the model's internal parameters, the process involves inferring how different triggers influence the model’s behavior and ensuring these triggers can effectively exploit any weaknesses.

To address these challenges, we propose an autoencoder-based approach for generating dynamic triggers to execute black-box backdoor attacks on IoT sensor applications. Unlike existing studies that primarily focus on images~\cite{gu2019badnets, liu2017neural}, our approach is novel in its application to time-series IoT data, focusing on minimizing perturbations on input. In doing so, our main contributions are as follows. %% SL: we can probably expand this paragraph

% %We perform backdoor attack on various sensor based HAR models. 
% In this paper, we demonstrate a novel and very effective trigger generation technique to perturb the input data to make the model misclassify the data to an incorrect activity class. Our trigger generation technique introduces very little perturbation, as shown in Figure \ref{perturbation-delta-genonly}, and such perturbation can be used to perform targeted backdoor attacks without retraining the classifier mode.
%using GAN based trigger generator model \cite{goodfellow2020generative} (Diffusion GAN ??).
%Our key contributions to this paper are:
\begin{itemize}
\item We propose a novel black-box trigger generation technique for injecting backdoor trigger patterns and facilitating targeted attacks into IoT-based applications such as gait recognition and human activity recognition.
Our black-box attack operates without modifying the classifier model and does not require full access to the training dataset or the model parameters. 

% We propose a novel black-box trigger generation technique that can perturb the sensor data to generate the trigger pattern that can be used to perform backdoor attacks on sensor-based HAR models.
% \item We perform a novel backdoor attack without modifying the classifier model; this attack can be performed in a complete black box setting and without having complete access to the training dataset used for training the classifier. 
% black-box approach; 

%\item We demonstrate that a backdoor trigger can be learned by a generator from a given pre-trained model without even having access to the training or test data.
% \item We define a new matric to measure the success of attack based on ASR and MAD.

\item We demonstrate two targeted attacks on three different sensor-based datasets and models and empirically prove the attack is successful in a black-box setting. In addition, we demonstrate the efficacy of our technique by comparing it against three baseline techniques and show that our approach achieves a higher attack success rate with minimal changes to the original input.
% using metrics such as MAE and MAPE.
%\item We also empirically prove that our generator-based trigger generation technique outperforms other baseline techniques in terms of ASR per MAD as shown in Figure~\ref{GAIT_Pertu_Base5} and ~\ref{perturbation-delta-genonly}.

\item We provide an extensive evaluation and robustness analysis of our technique by running it through state-of-the-art adversarial defense mechanisms such as adversarial training, neural pruning, and activation clustering. Our results show that these defenses are largely ineffective in detecting or mitigating the attack.

\end{itemize}

\section{Background}

\subsection{Sensor-based IoT Systems}
Sensor-based IoT systems leverage data collected from sensing devices, such as wearables and IoT devices, to offer insights into user activities. These devices typically use inertial measurement unit (IMU) sensors, including accelerometers and gyroscopes, to generate time-series data. This data captures dynamic patterns of movement and orientation, enabling the recognition of various activities such as walking, running, and gait analysis. 

While there are approaches to analyzing sensor data, recent years have seen a rise in deep learning-based methods in recognizing complex patterns and activities~\cite{chen2024computer}. Deep learning models, such as recurrent neural networks (RNNs) and convolutional neural networks (CNNs), are particularly effective in handling the temporal dependencies and high dimensionality of time-series data~\cite{mohammadi2024deep}.  Essentially, time-series data from various sensors is used to train deep learning algorithms to infer various activities. 

Similar to prior work~\cite{malekzadeh2017replacement}, we assume that users subscribe to online services for their sensor data analysis. In this architecture, users share their time-series sensor data with an external service provider for analysis. The sensor readings are transmitted from the device to the cloud, where they are analyzed using deep learning algorithms. For instance, users may wear a smartwatch that collects IMU data, which is then sent to the cloud for gait analysis to provide authentication services. We assume that the service provider utilizes deep learning algorithms to perform these analyses, ensuring accurate and efficient results.

% The presence of motion sensors such as accelerometers, gyroscopes, and magnetometers in mobile phones has significantly helped advance human activity recognition in various applications, such as health tracking.
% How it works

% What are the sensors in wearables?

% What are the HAR features

% Motion sensor data can be used to recognize human daily activities and for more advanced security applications, such as GAIT-based authentication.
% GAIT is a manner of limb movements made during a person's motion, such as walking. GAIT data analysis is an important part of the early detection of Parkinson's disease or post-stroke analysis 
% \cite{wang2024gait, jiao2024systematic}.

\subsection{Black-box Backdoor Attacks}
Backdoor attacks manipulate a deep learning model's behavior by introducing hidden triggers during training. The model performs normally during inference and produces correct outputs for clean input data. However, when specific triggers are present in the input data, the model is manipulated to produce incorrect or predefined outputs. The key challenge is generating poisoned inputs --- inputs containing triggers that would cause the model to misclassify. This is represented as follows:
\begin{equation}
 x'(t) = x(t) + \delta(t)
\end{equation}
where $x'(t)$ is the poisoned input at time $t$, $x$ is the clean input at time $t$, and $\delta(t)$ is the trigger -- perturbation to the signal at time $t$. For brevity, we omit $t$ in our subsequent notation.

In a black-box backdoor attack, the trigger $\delta$ is generated without knowledge of the training model or the data. The adversary cannot access the model's internal parameters or the training process. Instead, they can only observe the model's predictions for various inputs. This scenario is more realistic, as attackers often cannot directly influence the training process but can interact with and observe the model's outputs. 
A key challenge in this context is determining how to effectively perturb the sensor data. If the perturbations to the input are too large, the cloud service provider may easily detect them. Conversely, the attack may be less effective if the perturbations are too subtle.

{\bf Attacker Goals}. Attackers may carry out the attack in different ways.  One approach, known as \textit{all-to-one}, is for the attacker to select a specific target class. The backdoor trigger is designed to cause the model to misclassify any input containing the trigger as this predefined target class, regardless of the actual input. This approach ensures that the model consistently produces the attacker’s desired output for inputs with the trigger.

Another approach called \textit{all-to-all}, involves manipulating the model so that any input containing the trigger is misclassified into a class determined by a function of the original input’s true class. For instance, this function could map the true class to any one of several potential target classes, creating a range of possible misclassifications. Such attacks are typically designed to reduce the accuracy of the model.  In this paper, we explore both the \textit{all-to-one} and \textit{all-to-all} attack approaches, analyzing how backdoor triggers can be designed to exploit vulnerabilities in each approach.

% In sensor-based recognition tasks,  the input data consists of time-series signals from IMU sensors like accelerometers and gyroscopes. Designing effective triggers for such time-series data involves embedding perturbations that can influence the model's behavior while remaining covert within the dynamic and variable nature of sensor data.

% There are different types of backdoor attacks.
% Targeted attack - The attacker chooses a target class and performs the attack in such a way that the attacker's goal is to misclassify given data from a non-targeted class to a target class.

% Input aware approach

% GAN based

% Diffusion GAN?

\subsection{Threat Model}

In our threat model, we consider a black-box attack scenario, which is a more realistic approach for adversarial attacks on deep learning models. In this scenario, the attacker does not have access to the model parameters or the ability to modify the training data. However, we assume the attacker can interact with the service provider's model by observing its predictions for various inputs, which allows them to generate a backdoor trigger. The attacker typically conducts what is known as a clean-label attack. In these attacks, the attacker manipulates only the input data without altering the true labels associated with the data. The goal is to exploit the model’s behavior so that it causes misclassification while keeping the true label of the input data unchanged.

We also assume that the attacker can manipulate time-series sensor data before it is transmitted to the service provider for analysis. This includes transforming or altering raw sensor data, such as information from accelerometers or gyroscopes. The attacker introduces perturbations into the data to exploit vulnerabilities in the model. 

% In our threat model, we consider the attacker to have only limited (a certain portion of data) access to the training data used for training the classifier. An attacker can use the trained classifier to run certain attacker-crafted input data to calculate the loss on a pre-trained model. The attacker has no knowledge of the model network or architecture (Blackbox).
% Moreover, the attacker doesn't modify the class label of the perturbed input data.

\subsection{Problem statement}
Given a set of $n$ sensor readings $\vb{x} = \{x_1, x_2, \ldots, x_n\}$, the raw signal is divided into fixed-size segments. Each input $x_i$ is a time series, where $ x_i \in \mathbb{R}^{T \times d} $, where $T$ represents the length of the time series and $d$ represents the dimension of the sensor readings. Without loss of generality, we assume the dimensions represent data from multiple sensors (e.g., accelerometer, gyroscope), and each sensor may have one or more dimensions (e.g., x-axis, y-axis, z-axis). Thus, given an input $x_i$, the goal of the classifier $F$ is to predict the label $y_i$.

In contrast, the goal of an adversary is to perform a targeted attack using a backdoor trigger. Specifically, given an input $x_i$, the adversary generates a trigger $\delta_i$ such that when $\delta_i$ is added into $x_i$, the modified input $x_i' = x_i + \delta$ causes the classifier $F$ to misclassify $x_i'$ into the attacker's target class $y_{adv}$, regardless of its true class.  Furthermore, the adversary only has access to the output of $F$ and cannot directly observe or modify the model’s internal parameters or training process. In addition, the adversary aims to minimize the perturbation introduced to the original signal. 

\section{Trigger Generation Framework}
% In this section, we describe different methods we use for trigger generation to perform adversarial attacks to a classifier model.
% Before we explain our trigger generation design, lets understand the challenges of designing a trigger generator in a HAR domain.
% \subsection {Challenges}
% HAR models use accelerometer signals as input typically at lower sampling rate around 50 Hz. So adding new sensor value in a time series sensor data is not feasible since it will be easily detected by simple examination. Moreover in order to inject a trigger into a classifier, we need to ensure the classifier accuracy is not compromised. A common technique used in trigger generation is poisoning the input data by adding noise and used during training process without changing the label (clean label attack)~\cite{}. The amount of noise added to the original sensor data has to be within a small range and the deviation of such noise has to be minimum to evade detection.
\begin{figure}[t]
  \centering
  %\includegraphics[width=\linewidth]{figures/Architecture.png}
  %https://docs.google.com/drawings/d/1F0qS2r1AxKGlgdPhGjzAyC8fd-VNOHfcQ-C9u7ORUU4/edit
  \includegraphics[width=3.3in]{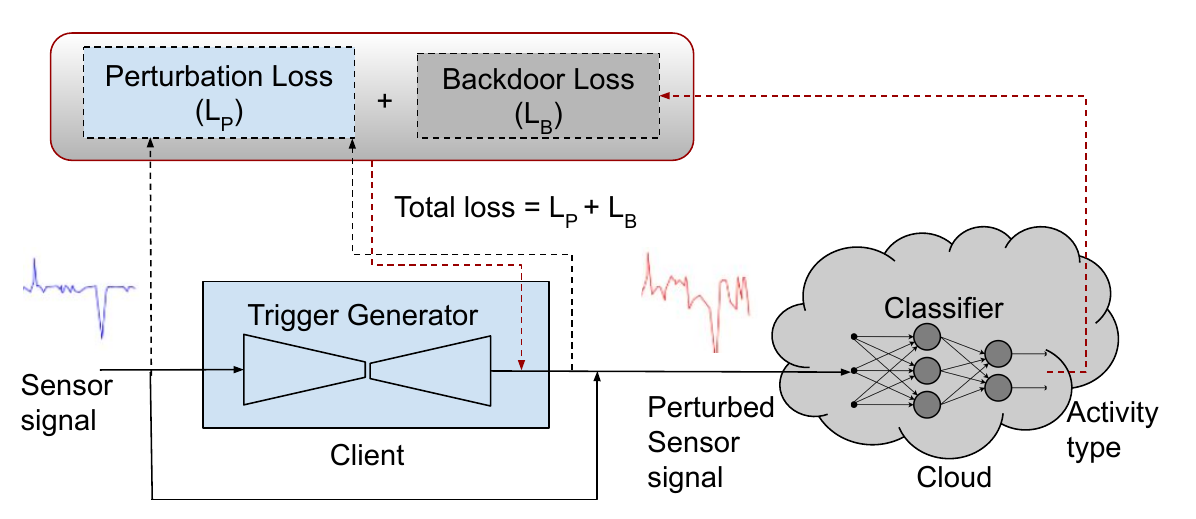}  
  \caption{Black-box backdoor attack architecture.}
  \label{fig:architecture}
  \vspace{-2mm}
\end{figure}
\subsection{Overview}
Our proposed framework is illustrated in Figure~\ref{fig:architecture}. The figure consists of two components: a black-box trigger generator and a DNN classifier. 
The black-box trigger generator runs on the client side and is based on an autoencoder framework~\cite{baldi2012autoencoders}. The autoencoder takes input sensor readings from various sensors and determines how to generate a trigger to manipulate the signal in such a way that the cloud model misclassifies it. Unlike traditional methods with a global trigger, our approach dynamically generates triggers for each input. The input signal itself is used to generate a unique trigger tailored to that specific input. The generated triggers are then added to the input. A key challenge we focus on is how to generate an effective trigger while minimizing the changes made to the raw input. After adding the trigger, the perturbed signal is sent to the cloud service.

The DNN classifier runs as a cloud service. It takes as input the poisoned input features $x'$ and outputs a prediction.  In our framework, we use the DNN's output to train the trigger generator. Specifically, the trigger generator is optimized to generate trigger $\delta$ that maximizes the likelihood of the DNN classifier misclassifying $x'$ into a target class $y_{adv}$.

\subsection{Black-box Trigger Generator}
We begin by discussing the traditional autoencoder and how we modify its basic structure to develop our black-box trigger generator. An autoencoder typically consists of two parts: an encoder $E$ and a decoder $D$. The encoder maps the input $x$ to a latent representation $z$, i.e., $z = E(x)$. The decoder then reconstructs the input from the latent representation $z$, producing $\hat{x} = D(z)$. The objective of a traditional autoencoder is to minimize the reconstruction loss $L(x, \hat{x})$ (e.g., mean squared error $L(x, \hat{x}) = ||x-\hat{x}||^2$), such that the prediction $\hat{x}$ closely approximates the original input $x$. Both the encoder and decoder may consist of one or more layers, enabling the autoencoder to learn complex data representations through this compression and reconstruction process.

To develop a black-box trigger generator, we modify the basic structure of the traditional autoencoder. Instead of focusing on reconstructing the input, we train to generate a trigger $\delta$ that causes the model to misclassify. 
This trigger $\delta$ is a perturbation applied to the original input $x' = x + \delta$. An alternative approach could involve directly generating the poisoned input with the trigger embedded rather than generating the trigger separately. However, our approach can also be used to implement the same concept.

Unlike the traditional autoencoder, which minimizes reconstruction loss, our black-box trigger generator uses a loss function consisting of two key components. First, because the trigger generator operates in a black-box setting, we cannot directly modify the internal workings of the target classifier or its decision-making process. However, we do have access to the output label for a given input, which allows us to determine whether the trigger successfully causes a misclassification.

Given the original input $x$, and the poisoned input $x'$, we want the target classifier $C$ to misclassify $x'$ to $y_{\text{adv}}$ defined by the attacker. Let $y'$ be the output label predicted by the classifier for the poisoned input $x'$. The backdoor loss corresponding to misclassification can be represented as:
\begin{equation}
L_{\text{B}}(y_{\text{adv}}, y') = \text{Loss}(y_{\text{adv}}, y')
\label{eqn:misclass}
\end{equation}
where $\text{Loss}(y_{\text{adv}}, y')$ is a cross-entropy loss to encourage the classifier to predict an incorrect label $y_{\text{adv}}$.

Second, we aim to keep the trigger $\delta$ as small as possible to make the perturbation minimal and potentially imperceptible. To achieve this, we introduce a regularization term on $\delta$. A common choice for this term is the $L_2$ norm of $\delta$:
\begin{equation}
L_{\text{P}} = ||\delta||^2    
\label{eqn:perturb}
\end{equation}
This regularization penalizes large perturbations, encouraging the generator to produce smaller triggers.

Combining (\ref{eqn:misclass}) and (\ref{eqn:perturb}), our total loss function is:
\begin{equation}
    L_{\text{total}} = L_{\text{B}}(y_{\text{adv}}, y') + \lambda L_{P}
    \label{eqn:loss}
\end{equation}
where $\lambda$ is a hyperparameter that controls the trade-off between causing misclassification and minimizing the size of the trigger $\delta$.

\begin{algorithm}[t]
\footnotesize
\caption{Trigger Generator training Algorithm.}
% Input: $x$ is the raw sensor training dataset; $X^{Test}$ is the test dataset. parameters $N$ represents the trigger data count, which is the total number of packets used for generator training; $C$ is the pre-trained classifier model; $L_{C}$ be the classifier loss; $G$ be the trigger generator; $L_{G}$ be generator loss; $L_{P}$ is the perturbation loss. $F$ is the number of features in the dataset. $b$ and $e$ are the number of batches and epochs used in the model training process. $\theta_{G}$ and $\theta_{C}$ are generator and classifier parameters respectively; 
% \textcolor{red}{$\alpha$ and $\beta$ are the weight parameters to tune the generator versus classifier loss.}
% \\
% Output: $X^{'}$ is the perturbed data after adding the trigger.}
\label{alg:generation_generator}
\begin{algorithmic}[1]
\Procedure{Train\_Generator}{}

% \State Init: $\theta_{G} \gets \Phi$  \Comment{initialize generator parameter}

% \State $X_{T} \gets \text{randomly chosen } N \text{ number of data points from } X$ \Comment{initialize Trigger training data}

%\State $\theta \gets \theta_t$
\For {$\text{epoch } i \gets 1,2,\cdots e$}
\For {$\text{batch } j \gets 1,2,\cdots n$}
\State $\delta^{(j)} \gets G(x^{(j)}; \theta_G)$ 
\State $x'^{(j)} \gets x^{(j)} + \delta^{(j)}$
\State $y'^{(j)} \gets C(x'^{(j)})$
\State $L_{\text{total}} \gets L_B + \lambda L_{P}$ \Comment{Using (\ref{eqn:misclass}) and (\ref{eqn:perturb})}
% \State $L_{\text{total}}.\text{backward}()$ \Comment{Compute gradients}
\State $\theta_{G} \gets \theta{_G} - \eta \nabla L_\text{total}(\theta_{G})$ \Comment{Compute gradients and update generator parameters $\theta{_G}$, $\eta$ is learning rate.}
% \For {$\text{each batch }i = 1,2,...b$}
%\State $L_{P} \gets \lvert D - G(D) \rvert$
% \State $L(\theta_{G};b) \gets \alpha L_{P}(G(X_{T}, \theta_{G}))+ \beta L_{C}(G(X_{T})+X_{T}, \theta_{C})$ \Comment{Compute loss to train generator}
 
% \State $\theta_{G} \gets \theta{_G} - \eta \nabla L(\theta_{G};b)$ \Comment{mini batch grad. descent}
\EndFor
\EndFor
\Return $G; \theta_G$

%\State $\triangle \theta_{t+1} = \theta - \theta_t$ \Comment{client local model update}
%\State $\zeta = \Vert \triangle \theta_{t+1} \Vert _2$ \Comment{norm of update}

%\State $X^{Trigger} \gets X+ G(X)$

%\Return $X^{Trigger}$

\EndProcedure
\end{algorithmic}
\end{algorithm}

Our trigger generation algorithm is shown in Algorithm~\ref{alg:generation_generator}.
 At each epoch, we perform the following steps. First, for a batch $j$ of input samples $x^{(j)}$, the trigger generator $G$ produces perturbed samples $\delta^{(j)} = G(x^{(j)}; \theta_G)$, where $\theta_G$ are the model parameters. We then use the target model to predict the output $y'^{(j)}$ for the poisoned input $x'^{(j)} = x^{(j)} + \delta^{(j)}$. We compute the backdoor loss $L_{\text{B}}$ by comparing these predicted outputs to the desired adversarial labels $y_{\text{adv}}$, as defined in (\ref{eqn:misclass}). And, we calculate the perturbation loss $L_{\text{P}}$ as defined in (\ref{eqn:perturb}).

We then combine the backdoor loss and the perturbation loss to obtain the final loss. Using this final loss, we perform backpropagation to compute the gradients with respect to the generator’s parameters. Finally, we update the generator’s parameters based on these gradients. This process repeats across epochs to iteratively train and refine the trigger generator.

\subsection{Attack Workflow}
Our backdoor attack operates in two stages: the training stage and the attack stage. In the training stage, the attacker trains the black-box trigger generator discussed above to learn how to generate the triggers. This requires the attacker to interact with the cloud service. That is,  it sends poisoned inputs and observes the corresponding outputs for training. The training process can be conducted offline, allowing the attacker to learn the trigger. 

In the attack stage, the attacker injects backdoor triggers into the input data before it is sent to the cloud service. This is done by intercepting the signal and inserting the trigger, which is designed to exploit vulnerabilities in the downstream model. For example, in a gait recognition system, the attacker could insert a carefully crafted trigger into the sensor data. This trigger could cause the model to misclassify any individual's gait as that of an authorized user, thereby allowing unauthorized access to secure systems or user accounts.

% \subsection{Implementation}
\section{Experimental Setup}
% We provide the detailed experimental setup in this section.
\subsection{Dataset}
We conduct experiments on three sensor datasets summarized in Table~\ref{tab:dataset}.
% 2. GAIT Authentication
% \begin{enumerate}
%     \item Dataset 5
%     \item Dataset 6
% \end{enumerate}
%In the GAIT authentication
\\
\textbf{Gait Authentication dataset}~\cite{zou2020gait} comprises 74,142 authentication samples collected from accelerometers and gyroscope sensors collected from 118 subjects. The sensor data was sampled at a rate of 50 Hz. For evaluation purposes, the training set includes data from 98 subjects, while the test set contains data from the remaining 20 subjects. Each authentication sample in the dataset is interpolated to a fixed length of 128 and consists of sensor data pairs, which can either be from two different subjects or from the same subject.
% horizontally hosted GAIT data of the same person or different person with two labels. 
% The dataset has 2 instances of 6 features with 128 times steps for the same person or different persons. 
% This GAIT authentication dataset can be used to judge whether a given pair of datasets belongs to the same person or not.
% The data was collected from 118 subjects. Among them, 20 subjects collected a larger amount of data in two days, with each having thousands of samples, and 98 subjects collected a smaller amount of data in one day, with each having hundreds of samples. Each data sample contains the 3-axis accelerometer data and the 3-axis gyroscope data. The sampling rate of all sensor data is 50 Hz.

% \begin{table}[t]
% \centering
% \caption{Summary of datasets used in our experiments.}
% \label{tab:dataset}
% \begin{tblr}{
%   row{even} = {c},
%   row{1} = {c},
%   row{3} = {c},
%   vline{-} = {1-4}{},
%   hline{1-5} = {-}{},
% }
% Dataset       & {Subject\\count} & {Sensors \\used}                 & {Number of\\classes} & {Number of\\features} \\
% Gait& 118                       & {accel., gyro.} & 2                    & 6                     \\
% Motion-sense& 24                        & {accel., gyro.} & 6                    & 12                    \\
% UCI & 30                        & {accel., gyro.} & 6                    & 9                     \\
% %              &                           &                                  &                      &                       
% \end{tblr}
% \end{table}

\begin{table}[t]
\centering
\caption{Summary of datasets used in our experiments.}
\label{tab:dataset}
\begin{tabular}{|c|c|c|c|c|}
\hline
Dataset      & \begin{tabular}[c]{@{}c@{}}Subject\\ count\end{tabular} & \begin{tabular}[c]{@{}c@{}}Sensors \\ used\end{tabular} & \begin{tabular}[c]{@{}c@{}}Number of\\ classes\end{tabular} & \begin{tabular}[c]{@{}c@{}}Number of\\ features\end{tabular} \\ \hline
Gait         & 118                                                     & accel., gyro.                                           & 2                                                           & 6                                                            \\ \hline
Motion-sense & 24                                                      & accel., gyro.                                           & 6                                                           & 12                                                           \\ \hline
UCI          & 30                                                      & accel., gyro.                                           & 6                                                           & 9                                                            \\ \hline
\end{tabular}
\end{table}

\textbf{Motion-sense dataset~\cite{Malekzadeh:2018:PSD:3195258.3195260, malekzadeh2019mobile}} consists of accelerometer and gyroscope measurements collected from 24 participants in 15 trials using an iPhone 6s. The sensor reading is sampled with a sample rate of 50Hz and corresponds to 6 activities: walking downstairs, walking upstairs, walking, jogging, sitting, and standing. The time-series data also has the personal attributes of participants: gender, age, weight, and height. There are a total of 139,873 samples --- 107,434 training and  32,439 test samples. Each sample has 12 features recorded over 50 time steps. 

\textbf{UCI HAR dataset~\cite{anguita2013public}} contains accelerometer and gyroscope data collected from 30 subjects using a smartphone for 6 activities. The sensor data is sampled at a frequency of 50 Hz and has 9 features over 128 time steps. The activities are standing, sitting, lying, walking, walking upstairs, and walking downstairs. In total, there are 10,299 samples in the dataset, with 7,352 training samples and 2,947 test samples.

% The sensor readings were filtered to remove noise and then segmented into windows of size 2.56 seconds (i.e. 128 readings per window) with 50\% overlap. The acceleration values were separated into gravity and body components using a Butterworth low-pass filter with a cutoff frequency of 0.3 Hz. From each window, 561 features were computed from the time and frequency domain. There are 10299 samples in the dataset, with 7352 training samples and 2947 test samples. 
%The availability of raw sensor reading in the dataset allows us to explore adversarial examples in both feature and signal domains.

%The data I used is the Human Activity Recognition dataset from the UCI Machine Learning Repository. It was created from experiments with 30 volunteers aged 19-48 performing 6 activities (walking, walking upstairs, walking downstairs, sitting, standing, laying) whilst wearing a Samsung Galaxy S2 on their waist. Tri-axial linear acceleration and Tri-axial angular velocity was captured at a constant rate of 50Hz, with pre-processing and calculations applied to create a dataset with 561 normalised and [-1,1] bounded features.

\subsection{Model}
In this section, we describe the various classification models that we target with backdoor attacks.

\textbf{Gait Authentication:} 
For Gait authentication, we utilize the network structure outlined in ~\cite{zou2020gait}. This architecture includes two CNN-based feature extractors, each processing a gait signal, followed by LSTM layers for capturing temporal dependencies and a linear layer for final classification. The model is trained using binary cross-entropy loss. For our trigger generator model, both the encoder and decoder consist of three linear layers with ReLU activations. The autoencoder is trained using Mean Squared Error (MSE).

% For GAIT authentication, the classifier model consists of 3 pairs of 2D convolution layers handling two portions of GAIT data in parallel, followed by 2 LSTM layers where the first LSTM takes the concatenated output of both convolution pipelines. The last layer is a linear layer that predicts binary classification. We use the Relu activation function and do batch normalization to avoid overfitting. We use cross-entropy as the loss function and Adam optimizer with a learning rate of 0.001.
% The Trigger Generator model consists of an encoder and decoder, each in one of three linear layers with Relu activations. We use Mean Squared Error (MSE) as the generator's loss function.

\textbf{Human Activity Recognition:} 
The classifier model consists of three 2D convolutional layers followed by a linear output layer. Each convolutional layer employs ReLU activations, and the dropout probability is set to 0.2. We use cross-entropy as the loss function and the Adam optimizer for training. This classification model is trained on both the UCI and MotionSense datasets. 
For the trigger generation model, we use the same network architecture as described above.

All models are implemented using the PyTorch framework, and we train them for 200 epochs with a batch size of 32.

% \textbf{Motion-sense:} 
% The classifier model consists of a series of 2D convolutional layers followed by a linear output layer. each layer uses Relu activations and a dropout probability of 0.2 to 0.4. We use cross-entropy as the loss function and Adam optimizer with a learning rate of 0.001.
% % The Trigger Generator model consists of an encoder and decoder, each in one of three linear layers with Relu activations and a dropout probability of 0.3. We use Mean Squared Error (MSE) as the generator's loss function.

% \textbf{UCI HAR:}
% For the UCI classifier model, we use a series of linear layers with Relu activations and a dropout of 0.3. We use cross-entropy as the loss function and Adam optimizer with a learning rate of 0.001.
% The Trigger Generator model consists of an encoder and decoder, each in one of three linear layers with Relu activations and a dropout probability of 0.3. We use Mean Squared Error (MSE) as the generator's loss function.

% \begin{enumerate}
%     \item CNN
%     \item LSTM
% \end{enumerate}
% Generator model
% Discriminator model

% Loss functions
% epoch
% batch

% Diffusions GAN ?

\subsection{Metrics}
We use the following metrics to measure the attack's performance and to assess the changes made to the sensor data.
%Quantify the perturbation 
% \begin{itemize}
    
    \textbf{Attack Success Rate (ASR)} is defined as the ratio of samples with triggers misclassified by the classifier to the total number of samples with triggers used in the attack.
% \begin{eqnarray*}
%   \text{A.S.R} =& \frac{\text{Number of samples with trigger successfully misclassified}}{\text{Total number of samples with trigger}}
% \end{eqnarray*}

    % \item \textbf{Confusion Matrix} evaluates the performance of the classification model and provides a detailed breakdown of the model's predictions compared to the actual ground truth across different classes. 

    \textbf{Mean Absolute Error (MAE)} provides the perturbation score across features and measures the overall change from the original sensor data. 
    It measures the average absolute difference between the original and backdoor data.  
    % taking the average of the sum of the absolute deviation of features of each sample. 
% \begin{eqnarray*}
%   \text{M.A.D} =& \frac{\text{absolute deviation of perturbation from mean perturbation}}{\text{Total number of samples in the attack}}
% \end{eqnarray*}

% \begin{eqnarray*}
%   \text{M.A.D} =& \frac{\text{absolute deviation of perturbation }}{\text{Total number of samples in the attack}}
% \end{eqnarray*}

% \begin{equation}
%     \text{M.A.D} = \frac{\sum_i \frac{\sum_j |\delta_{ij}|}{C} } {N}
% \end{equation}
% where $\delta$ is the perturbation of a datapoint, $x$ is the original datapoint, $N$ is the number of datapoints in the dataset, and $i$, $j$ are the row and column indices respectively. $C$ represents the number of features, and $N$ is the number of data points in the dataset.

    \textbf{Mean Absolute Percentage Error (MAPE)} 
    measures the percentage deviation of original and backdoor data. 
    % provides the mean of absolute perturbation percentage of the sum of deviations of features of each sample.
%entire test dataset.
% \begin{eqnarray*}
%   \text{M.A.P.D} =& \frac{\text{Mean absolute deviation of perturbation from original data}}{\text{Mean of original data}}
% \end{eqnarray*}

% \begin{equation}
%     \text{M.A.P.D} = \frac{\sum_j \frac{\sum_i |\delta_{ij}|}{\sum_i{x_{ij}} } } {N}
% \end{equation}

% \begin{equation}
%     \text{M.A.P.D} = \frac{\sum_i \frac{\sum_j |\delta_{ij}|}{\sum_j|{x_{ij}}| } } {N}
% \end{equation}
% where 
%  $\delta$ is the perturbation of a datapoint, $x$ is the original datapoint, $N$ is the number of datapoints in the dataset, and $i$, $j$ are the row and column indices respectively.
% If we interchange $i$ and $j$, we get datapoint-level MAPD instead of feature-level MAPD.
 
% \begin{enumerate}
%     \item ASR
%     \item Mean Absolute Deviation (MAD) = Mean of absolute deviation of perturbation from mean perturbation
%     \item Perturbation Power of generator = mean of L2 norm of successful adversarial attacks.
% \end{enumerate}
% \end{itemize}

\subsection{Baseline Techniques}
In order to evaluate the efficacy of our technique, we compare it against the following three baseline trigger-generation methods:
% \begin{itemize}
    % \item 

    \textbf{Random perturbation}: This technique involves randomly perturbing the data within a range of $(-k, k)$ to generate the trigger. 
    % This technique yields an average MAD of $k/2$, but MAPD varies depending on the magnitude of the original sensor data. 

    % \item
    \textbf{Fixed perturbation:} 
    In this method, we apply a fixed perturbation of $k$ when the original sensor value is positive; otherwise, when the original sensor value is negative, $-k$.
    %depending on the original signal direction. 
    The idea is to introduce fixed patterns into the data that can deceive the classifier into misclassifying the input.
    % We fix the perturbation to $k$ or $-k$, depending upon the sign of the sensor signal. This technique yields an average MAD of $k$, but the MAPD varies depending on the magnitude of the original sensor data.
    
    % \item 
    \textbf{Zero Masking:}
    In this technique, we set the sensor readings to zero for specific continuous time steps. The underlying idea is that muting specific time periods can act as a trigger pattern to mislead the classifier. 

\section{Results}

We conduct our experiments in various settings to demonstrate the effectiveness of our backdoor technique. All the classifier models we use in our experiments are pre-trained with original clean training data with high accuracy, as shown in the confusion matrices in Figure \ref{fig:GAIT_CM}(a) and \ref{fig:CM_Motionsense_all2all}(a).
%, and ~\ref{fig:CM_UCI_all2all}(a). 
Unless specified otherwise, we set the attack mode of \textit{all-to-one} as the default mode of attack in the experiments.
% The following subsections provide a detailed analysis of our experimental results and key takeaways.

\subsection{Baseline performance comparison} 
% \begin{table*}[t]
% \centering
% \caption{Attack performance comparison of different techniques on different datasets.}
% \label{tab:performance1}
% \begin{tblr}{
%   cells = {c},
%   cell{1}{1} = {c=2,r=2}{},
%   cell{1}{3} = {c=9}{},
%   cell{2}{3} = {c=3}{},
%   cell{2}{6} = {c=3}{},
%   cell{2}{9} = {c=3}{},
%   vlines,
%   hline{1,3-9} = {-}{},
%   hline{2} = {3-11}{},
% }
% {Attack\\ techniques} &                 & Dataset &      &           &              &      &           &      &      &           \\
%                       &                 & Gait    &      &           & Motion-sense &      &           & UCI  &      &           \\
% Attack mode           & Noise parameter & ASR     & MAE  & MAPE (\%) & ASR          & MAE  & MAPE (\%) & ASR  & MAE  & MAPE (\%) \\
% Random                & Uniform(-1, 1)  & 0.07    & 0.5  & 83        & 0.20         & 0.5  & 28        & 0.67 & 0.5  & 112       \\
% Fixed                 & \{-1, 1\}       & 0.39    & 1.0  & 104       & 0.72         & 1.0  & 50        & 0.67 & 1.0  & 472       \\
% Fixed                 & 5 or -5         & 0.84    & 5.0  & 289       & 0.81         & 5.0  & 260       & 0.72 & 5.0  & 1700      \\
% Zero masking           & 50/5/3          & 0.14    & 1.92 & 123       & 0.03         & 0.61 & 77        & 0.02 & 0.28 & 84        \\
% Ours                  & NA              & 0.91    & 0.22 & 12        & 0.98         & 0.08 & 17        & 0.97 & 0.09 & 26        
% \end{tblr}
% \end{table*}

% \usepackage{array}
% \usepackage{multirow}
\begin{table*}[t]
\centering
\caption{Attack performance comparison of different techniques on different datasets.}
\label{tab:performance1}
\begin{tabular}{|cc|ccccccccc|}
\hline
\multicolumn{2}{|c|}{\multirow{2}{*}{\begin{tabular}[c]{@{}c@{}}Attack\\ Techniques\end{tabular}}} & \multicolumn{9}{c|}{Dataset}                                                                                                                                                                                                                        \\ \cline{3-11} 
\multicolumn{2}{|c|}{}                                                                             & \multicolumn{3}{c|}{Gait}                                                              & \multicolumn{3}{c|}{Motion-sense}                                                      & \multicolumn{3}{c|}{UCI}                                          \\ \hline
\multicolumn{1}{|c|}{Attack mode}                         & Noise parameter                        & \multicolumn{1}{c|}{ASR}  & \multicolumn{1}{c|}{MAE}  & \multicolumn{1}{c|}{MAPE (\%)} & \multicolumn{1}{c|}{ASR}  & \multicolumn{1}{c|}{MAE}  & \multicolumn{1}{c|}{MAPE (\%)} & \multicolumn{1}{c|}{ASR}  & \multicolumn{1}{c|}{MAE}  & MAPE (\%) \\ \hline
\multicolumn{1}{|c|}{Random}                              & Uniform(-1, 1)                         & \multicolumn{1}{c|}{0.07} & \multicolumn{1}{c|}{0.5}  & \multicolumn{1}{c|}{83}        & \multicolumn{1}{c|}{0.20} & \multicolumn{1}{c|}{0.5}  & \multicolumn{1}{c|}{28}        & \multicolumn{1}{c|}{0.67} & \multicolumn{1}{c|}{0.5}  & 112       \\ \hline
\multicolumn{1}{|c|}{Fixed}                               & \{-1, 1\}                              & \multicolumn{1}{c|}{0.39} & \multicolumn{1}{c|}{1.0}  & \multicolumn{1}{c|}{104}       & \multicolumn{1}{c|}{0.72} & \multicolumn{1}{c|}{1.0}  & \multicolumn{1}{c|}{50}        & \multicolumn{1}{c|}{0.67} & \multicolumn{1}{c|}{1.0}  & 472       \\ \hline
\multicolumn{1}{|c|}{Fixed}                               & \{5, -5\}                                & \multicolumn{1}{c|}{0.84} & \multicolumn{1}{c|}{5.0}  & \multicolumn{1}{c|}{289}       & \multicolumn{1}{c|}{0.81} & \multicolumn{1}{c|}{5.0}  & \multicolumn{1}{c|}{260}       & \multicolumn{1}{c|}{0.72} & \multicolumn{1}{c|}{5.0}  & 1700      \\ \hline
\multicolumn{1}{|c|}{Zero masking}                        & 50/5/3                                 & \multicolumn{1}{c|}{0.14} & \multicolumn{1}{c|}{1.92} & \multicolumn{1}{c|}{123}       & \multicolumn{1}{c|}{0.03} & \multicolumn{1}{c|}{0.61} & \multicolumn{1}{c|}{77}        & \multicolumn{1}{c|}{0.02} & \multicolumn{1}{c|}{0.28} & 84        \\ \hline
\multicolumn{1}{|c|}{Ours}                                & NA                                     & \multicolumn{1}{c|}{0.91} & \multicolumn{1}{c|}{0.22} & \multicolumn{1}{c|}{12}        & \multicolumn{1}{c|}{0.98} & \multicolumn{1}{c|}{0.08} & \multicolumn{1}{c|}{17}        & \multicolumn{1}{c|}{0.97} & \multicolumn{1}{c|}{0.09} & 26        \\ \hline
\end{tabular}
\end{table*}

We start by comparing our attack technique with baseline perturbation methods across three datasets: Gait, MotionSense, and UCI. We set the hyperparameters for our baseline algorithms as follows. For the random perturbation technique, we generate a uniformly random perturbation within the range of (-1, 1) and add it to the original sensor data, keeping the signal alterations minimal. In the fixed perturbation technique, we apply a perturbation of 1 or -1 based on the sign of the sensor signal and also evaluate higher perturbation values of 5 or -5. For zero masking, we set every third feature value to zero for five adjacent samples in every 50th sample; we represent the parameter as (50/5/3). 

Table \ref{tab:performance1} presents the comparative performance of these techniques across each dataset. For the random perturbation technique, although the absolute perturbation is low at 0.5, the percentage change remains high at 83\% for the Gait dataset. Despite this significant percentage change, the attack success rate remains very low, indicating that randomly generating triggers are ineffective for successful backdoor exploits. Similarly, the attack success rate is only 0.2 for the MotionSense dataset and 0.67 for the UCI dataset, even though the perturbation levels are still high, particularly for the UCI dataset, where the percentage change is 112\%.

\begin{figure}[t]
  \centering
  \includegraphics[width=2.8in]
  %{figures/BaselineFixed5_perturbation.png}
  %{figures/GAIT_fixed5_1_fig_full.pdf} %GAIT_local_1_fig_full_wide_fixed5
  %GAIT_local_new_1_fig_full_wide.pdf
  %{figures/GAIT_local_1_fig_full_wide_fixed5.pdf}
  {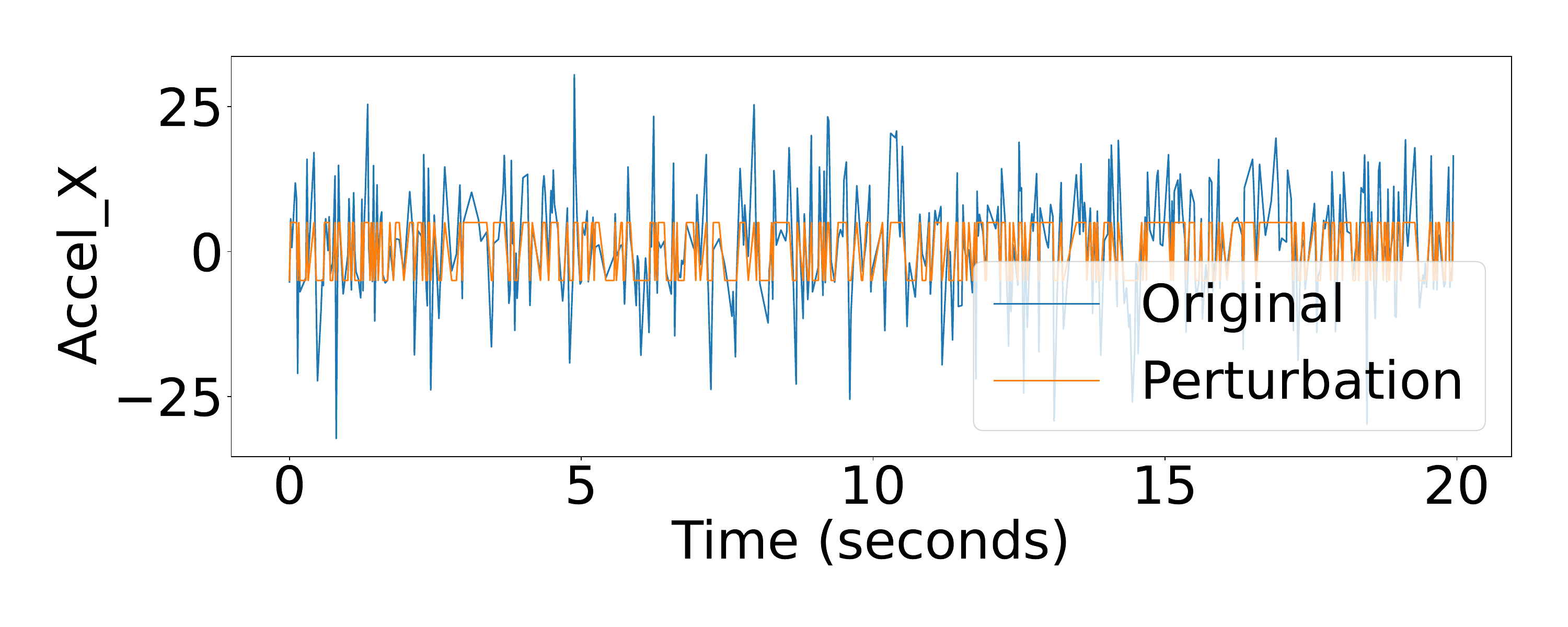}
  \caption{Perturbation(delta) generated corresponding to accelerometer X-axis using baseline technique with fixed perturbation of 5 and -5 (MAE = 5 and MAPE of 289\%), giving ASR of 0.84 on the Gait dataset.}
  %\Description{Perturbation - Fixed.}
  \label{GAIT_Pertu_Base5}
\end{figure}

For the fixed perturbation technique, although the attack success rate is 0.39, which is higher than the random perturbation method, the magnitude of the perturbation is also greater. This indicates that increasing the perturbation can lead to higher misclassification rates. When we increase the fixed perturbation to 5 and -5, the attack success rate (ASR) rises significantly to 0.84. However, this level of perturbation is easily detectable visually, as shown in Figure~\ref{GAIT_Pertu_Base5}. We observe a similar trend in both the UCI and Motion-sense datasets.

The zero masking technique also demonstrates a relatively low attack success rate. Despite setting certain values to zero, this method still introduces a significant level of perturbation because the zeroed values can create abrupt changes in the data sequence, which are not typically present in the original signal. We also experimented with various other settings for this technique, which are not shown in the table, but these adjustments also resulted in a low attack success rate. This suggests that simply masking data to zero, even with different configurations, is not sufficient to achieve a high success rate in backdoor attacks. 

Compared to all baseline techniques, our method achieves a high attack success rate with minimal perturbation, as reflected by the low percentage change in the data. Specifically, our approach yields attack success rates of 0.91 for the Gait dataset, 0.98 for the MotionSense dataset, and 0.97 for the UCI dataset. This demonstrates that our technique not only maintains a high level of effectiveness across various datasets but also does so with significantly lower perturbation compared to other methods.

{\bf Takeaway:} \textit{Generator-based trigger generation is more effective than baseline techniques because it systematically learns to fool the classifier by optimizing both generator and classifier losses. In contrast, baseline techniques need much higher perturbations to achieve similar attack success rates.}

\subsection{Attack Scenarios}
\begin{table*}[t]
\centering
\footnotesize
\caption{Attack performance of our technique on different attack types.}% (take from table 2)
\label{tab:performanceall2one}
\begin{tabular}{|cc|ccccccccc|}
\hline
\multicolumn{2}{|c|}{\multirow{2}{*}{Attack type}}         & \multicolumn{9}{c|}{Dataset}                                                                                                                                                                                                                        \\ \cline{3-11} 
\multicolumn{2}{|c|}{}                                     & \multicolumn{3}{c|}{Gait}                                                              & \multicolumn{3}{c|}{Motion-sense}                                                      & \multicolumn{3}{c|}{UCI}                                          \\ \hline
\multicolumn{1}{|c|}{Attack mode}                 & Target & \multicolumn{1}{c|}{ASR}  & \multicolumn{1}{c|}{MAE}  & \multicolumn{1}{c|}{MAPE (\%)} & \multicolumn{1}{c|}{ASR}  & \multicolumn{1}{c|}{MAE}  & \multicolumn{1}{c|}{MAPE (\%)} & \multicolumn{1}{c|}{ASR}  & \multicolumn{1}{c|}{MAE}  & MAPE (\%) \\ \hline
\multicolumn{1}{|c|}{\multirow{6}{*}{All-to-one}} & 0      & \multicolumn{1}{c|}{0.91} & \multicolumn{1}{c|}{0.22} & \multicolumn{1}{c|}{12}        & \multicolumn{1}{c|}{0.98} & \multicolumn{1}{c|}{0.08} & \multicolumn{1}{c|}{17}        & \multicolumn{1}{c|}{0.97} & \multicolumn{1}{c|}{0.09} & 26        \\ \cline{2-11} 
\multicolumn{1}{|c|}{}                            & 1      & \multicolumn{1}{c|}{0.90} & \multicolumn{1}{c|}{0.20} & \multicolumn{1}{c|}{12}        & \multicolumn{1}{c|}{0.97} & \multicolumn{1}{c|}{0.19} & \multicolumn{1}{c|}{17}        & \multicolumn{1}{c|}{0.96} & \multicolumn{1}{c|}{0.12} & 29        \\ \cline{2-11} 
\multicolumn{1}{|c|}{}                            & 2      & \multicolumn{1}{c|}{NA}   & \multicolumn{1}{c|}{NA}   & \multicolumn{1}{c|}{NA}        & \multicolumn{1}{c|}{0.96} & \multicolumn{1}{c|}{0.18} & \multicolumn{1}{c|}{19}        & \multicolumn{1}{c|}{0.95} & \multicolumn{1}{c|}{0.11} & 28        \\ \cline{2-11} 
\multicolumn{1}{|c|}{}                            & 3      & \multicolumn{1}{c|}{NA}   & \multicolumn{1}{c|}{NA}   & \multicolumn{1}{c|}{NA}        & \multicolumn{1}{c|}{0.97} & \multicolumn{1}{c|}{0.19} & \multicolumn{1}{c|}{19}        & \multicolumn{1}{c|}{0.96} & \multicolumn{1}{c|}{0.13} & 29        \\ \cline{2-11} 
\multicolumn{1}{|c|}{}                            & 4      & \multicolumn{1}{c|}{NA}   & \multicolumn{1}{c|}{NA}   & \multicolumn{1}{c|}{NA}        & \multicolumn{1}{c|}{0.99} & \multicolumn{1}{c|}{0.20} & \multicolumn{1}{c|}{22}        & \multicolumn{1}{c|}{0.96} & \multicolumn{1}{c|}{0.12} & 27        \\ \cline{2-11} 
\multicolumn{1}{|c|}{}                            & 5      & \multicolumn{1}{c|}{NA}   & \multicolumn{1}{c|}{NA}   & \multicolumn{1}{c|}{NA}        & \multicolumn{1}{c|}{0.99} & \multicolumn{1}{c|}{0.19} & \multicolumn{1}{c|}{19}        & \multicolumn{1}{c|}{0.95} & \multicolumn{1}{c|}{0.11} & 25        \\ \hline
\multicolumn{1}{|c|}{All-to-all}                  & NA     & \multicolumn{1}{c|}{0.93} & \multicolumn{1}{c|}{0.25} & \multicolumn{1}{c|}{13}        & \multicolumn{1}{c|}{0.95} & \multicolumn{1}{c|}{0.21} & \multicolumn{1}{c|}{24}        & \multicolumn{1}{c|}{0.93} & \multicolumn{1}{c|}{0.15} & 31        \\ \hline
\end{tabular}
\end{table*}

%We now evaluate our technique on two target attack scenarios.
We now evaluate our technique in two target attack scenarios. In this experiment, we set the trigger data percentage --- the proportion of training data used to train our generator model --- to 70\%. Although we assume the attacker has access to the original training data, we will later demonstrate that the attack can still be successful even if the attacker’s dataset is disjoint from the training data.

% \begin{figure}[t]
%   \centering
%   \includegraphics[width=2.8in]%{figures/Generatoronly_pdelta.png}
%   %{figures/GAIT_ours_1_fig_full.pdf}
%  {figures/GAIT_local_1_fig_full_wide.pdf}
%  \caption{Comparison of perturbation values generated by our technique with the original signal values.}
%   % \caption{With very little perturbation (MAE = 0.2 and MAPE of 12\%) on Gait dataset using our trigger generation method giving ASR of 0.91.}
%   %\Description{Perturbation - Delta.}
%   % \vspace{-5mm}
%   \label{perturbation-delta-genonly}
% \end{figure}

Recall that in the \textit{All-to-One Target Attack}, the attacker selects a target label, and the attacker’s goal is to generate a trigger that misclassifies a given input sample as the chosen target label. Table~\ref{tab:performanceall2one} shows the attack performance across various target classes for \textit{all-to-one} target attacks. Our technique achieves a high attack success rate, regardless of the chosen target label, with rates exceeding 0.9 for the Gait dataset and 0.95 for the MotionSense and UCI datasets. Additionally, the perturbation introduced by our trigger generation technique is minimal. 
% Figure~\ref{perturbation-delta-genonly} illustrates that the perturbation values are close to 0.2, indicating only slight modifications to the signal.

\begin{figure}[t]
    \centering
    \subfigure[Clean data]{\includegraphics[width=0.24\textwidth]{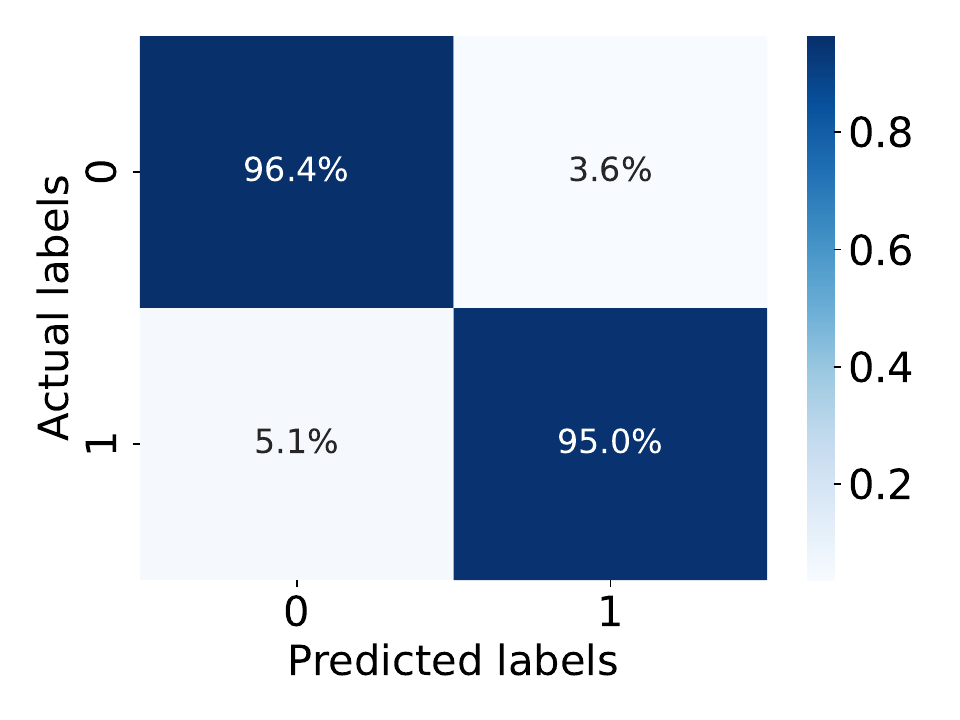}}
    \subfigure[Backdoor data]{\includegraphics[width=0.24\textwidth]{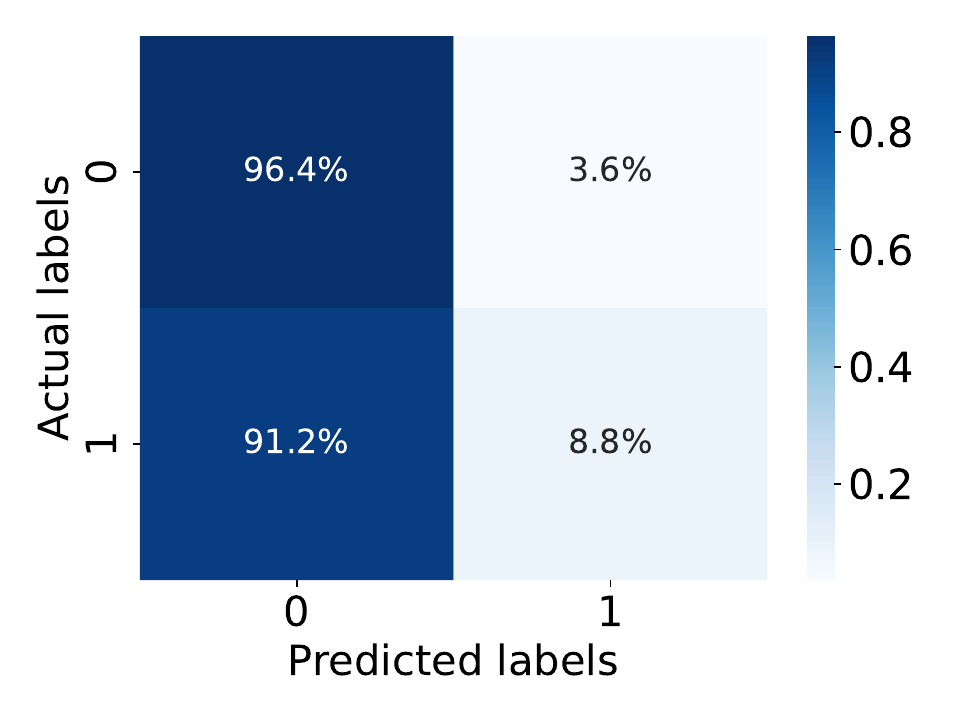}}
    \caption{All-to-all attack confusion matrix on Gait dataset.}
    % (a) Clean data on the clean model, (b) Adversarial data generated using our trigger generator on the clean model in case of Gait dataset with a train-test ratio of 80-20 and 70\% of training data is used for trigger generator training.}
    \label{fig:GAIT_CM}
    \end{figure}

    \begin{figure}[t]
    \centering
    \subfigure[Clean data]{\includegraphics[width=0.24\textwidth]%{figures/Motionsense_CM_Clean.png}} %CM_Multitask_Motionsense_Clean
    {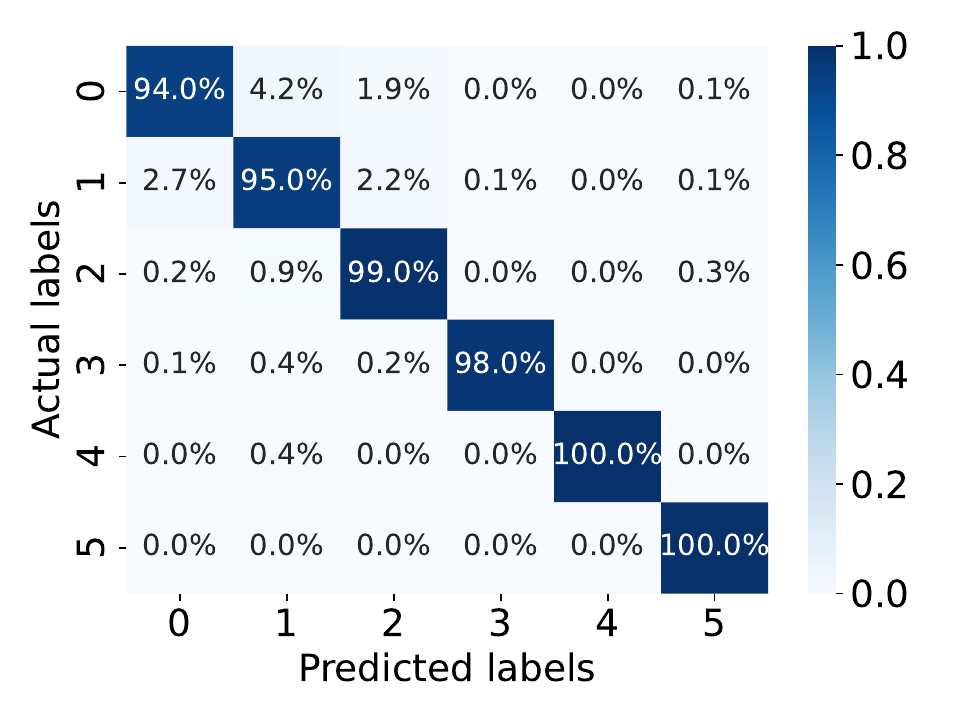}}
    \subfigure[Backdoor data]{\includegraphics[width=0.24\textwidth]%{figures/MotionsenseAll2All.png}} %CM_Multitask_MotionsenseAll2All
    {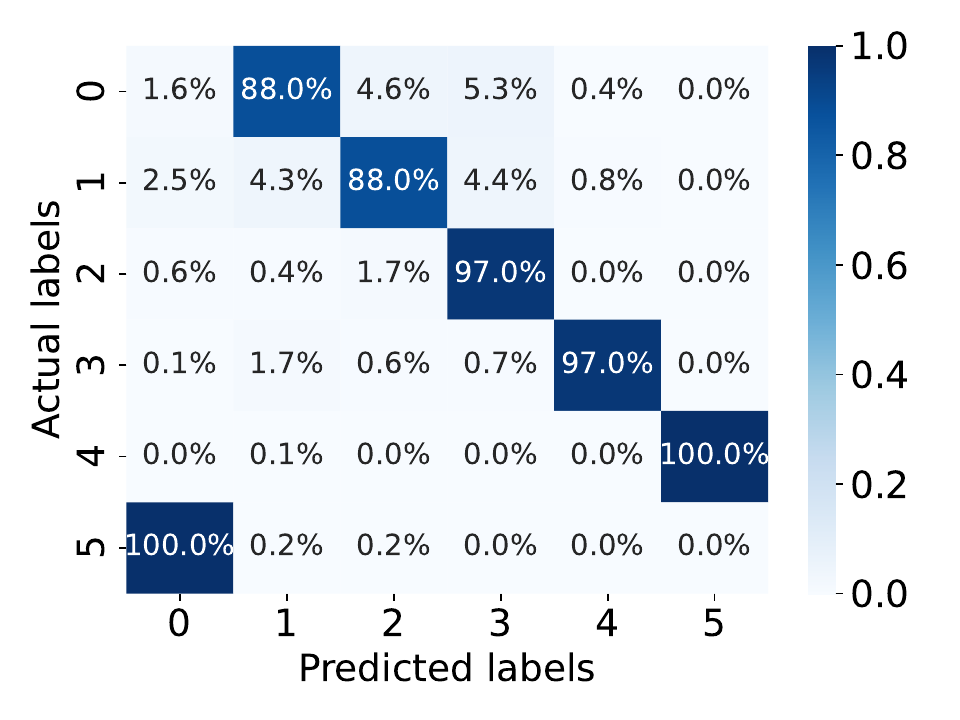}}
    \caption{All-to-all attack confusion matrix on Motion-sense dataset.}
    % in case of (a) clean data on the clean model, (b) \textit{all-to-all} attack on Motion-sense dataset.}
    \label{fig:CM_Motionsense_all2all}
    \vspace{-2mm}
    %\Description{Whole dataset is combined and split randomly into training and test}
\end{figure}

In the \textit{All-to-All Target Attack}, the attacker aims to misclassify each class to the next target label; for example, label 0 is classified as 1, label 1 as 2, and so on. In the Gait dataset, which has two labels, this means misclassifying label 0 as 1 and label 1 as 0. Table~\ref{tab:performanceall2one} demonstrates the effectiveness of our approach, showing that we achieve a high attack success rate while maintaining low perturbation changes. This is further evidenced by the confusion matrices for the Gait and UCI datasets. As shown in Figure~\ref{fig:GAIT_CM} and Figure~\ref{fig:CM_Motionsense_all2all}, while the model performs accurately on clean data, it misclassifies samples to the next target label when presented with backdoored data.

\begin{figure*}[ht]
  \centering
  \includegraphics[width=7in]
  %{figures/BaselineFixed5_perturbation.png}
  %{figures/UCI_ours_5_fig_combined.pdf}
  %{figures/UCI_ours_local_5_fig_combined_new.pdf}
  {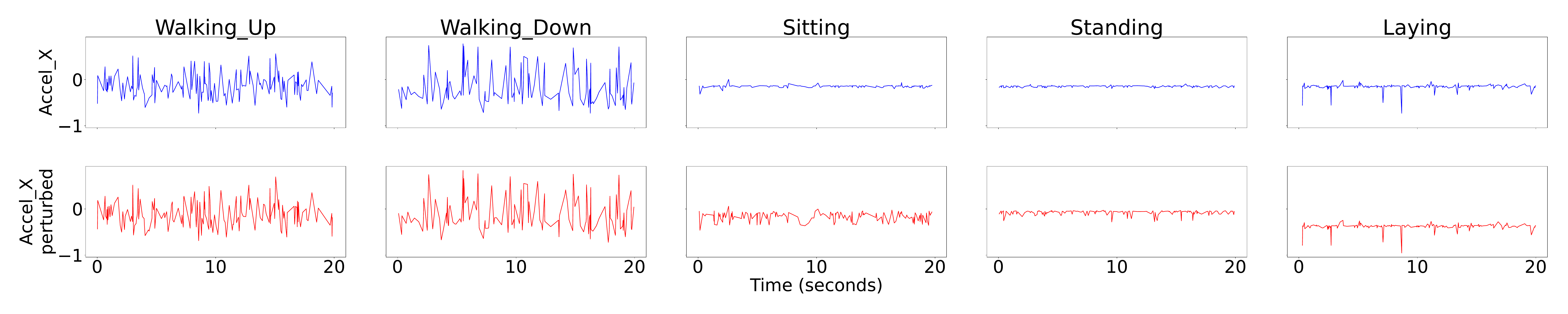}
  \caption{Comparison of original and backdoor accelerometer readings on the UCI dataset. The top row indicates the original data corresponding to different activities, while the bottom row shows the corresponding backdoor data (i.e., original + trigger).}
  %\Description{Perturbation - Fixed.}
  \label{UCI_ours}
\end{figure*}

\begin{figure}[t]
  \centering
  \includegraphics[width=2.8in]%{figures/Generatoronly_pdelta.png}
    %{figures/GAIT_delta_1_fig_perturbation.pdf}
    {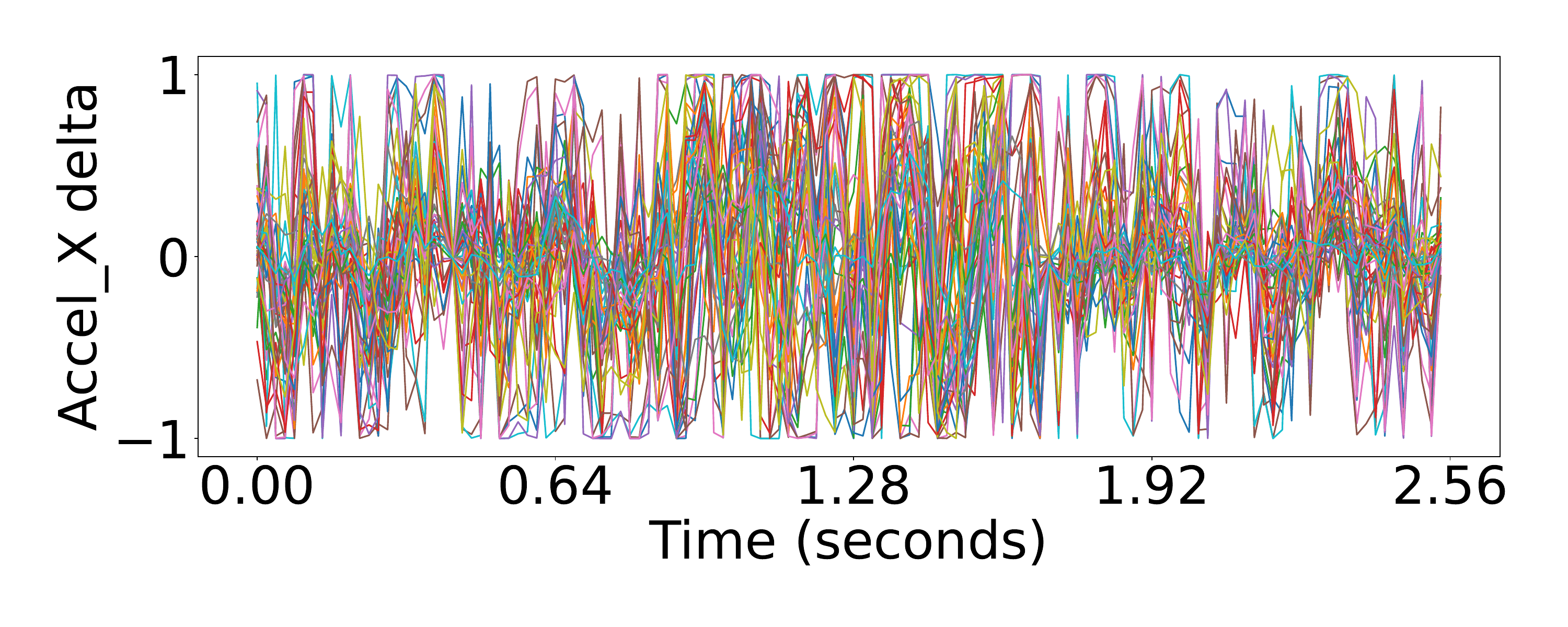}
    \vspace{-2mm}
    \caption{Unique trigger generated for different input samples on Gait dataset.}
  % \caption{Perturbation (delta) corresponding to first 50 input sensor data in case of  Gait dataset showing the uniqueness of perturbations per input sensor data.}
  %\Description{Perturbation - Delta.}
  % \vspace{-5mm}
  \label{perturbation-delta-genonly_GAIT}
\end{figure}

We now analyze the various triggers generated by our trigger generator. Figure~\ref{UCI_ours} illustrates the triggers created for different activities. As shown, the backdoored data closely resembles the original signal, indicating that not only is the perturbation minimal, but the changes do not significantly alter the underlying pattern. Additionally, we plot the various perturbations generated for the Gait dataset, as shown in Figure~\ref{perturbation-delta-genonly_GAIT}. We observe that each perturbation is unique and tailored to the specific input signal, highlighting the stealthiness of the attack. This uniqueness suggests that defense techniques that use fixed trigger patterns may not effectively detect the backdoor, as the triggers adapt to the data and maintain subtlety.

{\bf Takeaway:} \textit{Our generator-based trigger generation achieves a high attack success rate across various target-based attack scenarios while maintaining very low perturbation.}

\begin{figure}[ht]
    \centering
    \vspace{-3mm}
    %\hspace{-8mm}
    \subfigure[attack success]{\includegraphics[width=0.24\textwidth]{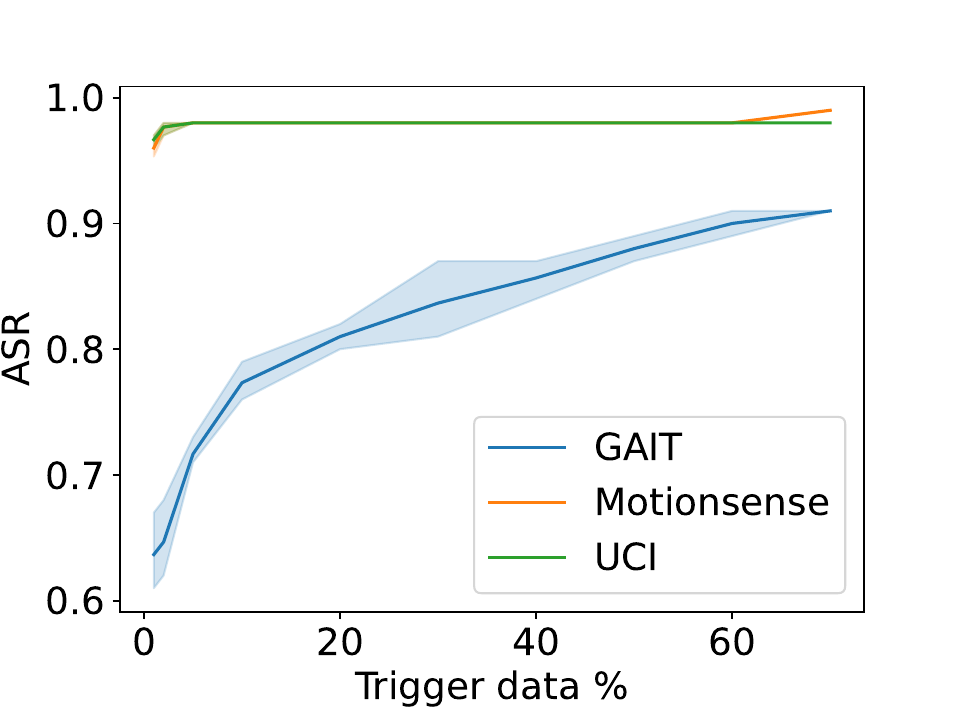}}
    % \hspace{-6mm}
    \subfigure[perturbation]{\includegraphics[width=0.24\textwidth]{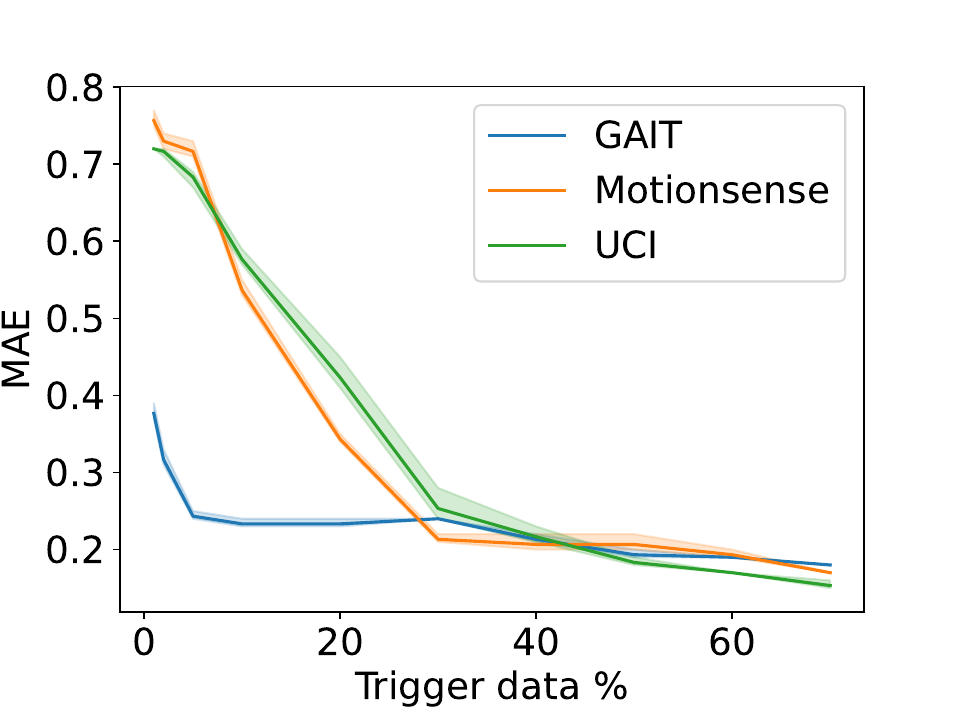}}
    %\subfigure[]{\includegraphics[width=0.32\textwidth]{figures/GAIT_CM_GENDIsc.png}} 
    % \vspace{-2mm}
    \caption{Varying backdoor trigger data \% for different datasets. }
    %, and (c) Adversarial data on the clean model 
    \label{fig:BD_ASR_MAD}
    %\Description{Whole dataset is combined and split randomly into training and test}
    \vspace{-3mm}
\end{figure}

\subsection{Impact of trigger data percentage}

Next, we examine how varying the percentage of backdoor trigger data—representing the proportion of training data used to train our generator—affects the attack’s effectiveness. Figure~\ref{fig:BD_ASR_MAD} presents the attack success rate and perturbation change for each dataset. As shown in Figure~\ref{fig:BD_ASR_MAD} (a), our technique remains effective even with a small amount of training data, as low as 2\%, for the UCI and MotionSense datasets. However, for the Gait dataset, we find that a larger proportion of training data is necessary, likely due to the uniqueness and variability of individual signals in this dataset. Specifically, the attack’s effectiveness improves with increased training data, indicating that having more data enables the generator to create more effective triggers (see Figure~\ref{fig:BD_ASR_MAD} (b)). Additionally, we observe that the required perturbation decreases as the amount of training data increases. This reduction in perturbation is likely because the generator becomes more adept at exploiting the classifier's weaknesses with more comprehensive training data.

{\bf Takeaway:} \textit{The amount of training data required for a successful attack can vary depending on the scenario. Generally, having access to more training data enhances the attacker's ability to execute a successful attack. Additionally, with more training data, the attack becomes more stealthy, as indicated by the smaller perturbation needed to achieve the desired outcome.}

\subsection{Disjoint trigger dataset scenario}
We now explore the scenario where the dataset used to train the generator is different from the dataset used to train the classifier. This represents a more realistic scenario, as the attacker may not have access to the training data.  To do this, we split our dataset into three parts: one for training the classifier, one for testing the classifier, and one for training the generator. Our goal is to assess whether the generator can be trained effectively without using data from the classifier's training set and to evaluate the attack's performance. We vary the percentage of data used for training the generator while keeping the test dataset fixed at 20\% of the overall dataset. Thus, a 10\% disjoint dataset indicates that we use the remaining 70\% for training the classifier and 20\% for evaluating our technique.

\begin{table}[t]
\centering
\caption{Impact of using the disjoint dataset for training the generator.}
\label{tab:DisjointPerformance}
\begin{tabular}{|c|cccccc|}
\hline
\multirow{3}{*}{\begin{tabular}[c]{@{}c@{}}Disjoint \\ trigger training\\ \%\end{tabular}} & \multicolumn{6}{c|}{Dataset}                                                                                                                     \\ \cline{2-7} 
                                                                                           & \multicolumn{2}{c|}{Gait}                             & \multicolumn{2}{c|}{Motion-sense}                     & \multicolumn{2}{c|}{UCI}         \\ \cline{2-7} 
                                                                                           & \multicolumn{1}{c|}{ASR}  & \multicolumn{1}{c|}{MAE}  & \multicolumn{1}{c|}{ASR}  & \multicolumn{1}{c|}{MAE}  & \multicolumn{1}{c|}{ASR}  & MAE  \\ \hline
70                                                                                         & \multicolumn{1}{c|}{0.81} & \multicolumn{1}{c|}{0.39} & \multicolumn{1}{c|}{0.98} & \multicolumn{1}{c|}{0.53} & \multicolumn{1}{c|}{0.98} & 0.56 \\ \hline
60                                                                                         & \multicolumn{1}{c|}{0.77} & \multicolumn{1}{c|}{0.45} & \multicolumn{1}{c|}{0.98} & \multicolumn{1}{c|}{0.56} & \multicolumn{1}{c|}{0.98} & 0.57 \\ \hline
50                                                                                         & \multicolumn{1}{c|}{0.74} & \multicolumn{1}{c|}{0.47} & \multicolumn{1}{c|}{0.98} & \multicolumn{1}{c|}{0.61} & \multicolumn{1}{c|}{0.98} & 0.61 \\ \hline
40                                                                                         & \multicolumn{1}{c|}{0.68} & \multicolumn{1}{c|}{0.48} & \multicolumn{1}{c|}{0.97} & \multicolumn{1}{c|}{0.63} & \multicolumn{1}{c|}{0.98} & 0.64 \\ \hline
30                                                                                         & \multicolumn{1}{c|}{0.64} & \multicolumn{1}{c|}{0.50} & \multicolumn{1}{c|}{0.97} & \multicolumn{1}{c|}{0.67} & \multicolumn{1}{c|}{0.97} & 0.66 \\ \hline
20                                                                                         & \multicolumn{1}{c|}{0.59} & \multicolumn{1}{c|}{0.51} & \multicolumn{1}{c|}{0.97} & \multicolumn{1}{c|}{0.74} & \multicolumn{1}{c|}{0.97} & 0.72 \\ \hline
10                                                                                         & \multicolumn{1}{c|}{0.54} & \multicolumn{1}{c|}{0.52} & \multicolumn{1}{c|}{0.95} & \multicolumn{1}{c|}{0.77} & \multicolumn{1}{c|}{0.96} & 0.74 \\ \hline
\end{tabular}
\end{table}
We observe that our attack achieves a high success rate for activity recognition models even when the training dataset for the generator differs from the classifier’s training data. Specifically, on the MotionSense and UCI datasets, the attack success rates are 0.95 and 0.97, respectively, even with 10\% disjoint data. This demonstrates that the attack remains effective despite the differences between the generator and classifier training datasets. However, the Gait dataset shows that a larger portion of disjoint data is necessary for a successful attack. Here, we require 70\% of the disjoint dataset to achieve an attack success rate of 0.81. Our analysis of the perturbations reveals that the overall changes introduced are minimal. As the amount of disjoint data increases, the generator can produce smaller perturbations while still effectively executing the attack.

%
% In the Motion-sense dataset, we get a high ASR of 0.598 even with a very low disjoint dataset of 10\% of the actual training data used for generator training. The corresponding MAE is 0.77, which is also low. Similarly, for the UCI dataset, with 10\% of the disjoint dataset, the attack works with an ASR of 0.96 with a MAE of 0.74. Again, the trend of increasing ASR with reducing MAE is observed with disjoint dataset scenarios as well.

% The confusion matrix in Figure~\ref{fig:GAIT_CM_70_20_10} shows the comparison between clean model accuracy and one with trigger generated using the generator-based technique in case of disjoint GAIT dataset.
% \begin{figure}[ht]
%     \centering
%     \subfigure[]{\includegraphics[width=0.24\textwidth]{figures/GAIT_CM_70_20_10bd_clean.png}}
%     \subfigure[]{\includegraphics[width=0.24\textwidth]{figures/GAIT_CM_70_20_10_Generator.png}}
%     %\subfigure[]{\includegraphics[width=0.32\textwidth]{figures/GAIT_CM_70_20_10_GenDisci.png}} 
%     \caption{Confusion matrix in case of (a) clean data on the clean model, (b) adversarial data generated using the trigger generator on the clean model in case of GAIT dataset with 70\% trigger training data, 20\% of test and 10\%  train data is used for Generator training.}
%     %, and (c) Adversarial data on the clean model
%     \label{fig:GAIT_CM_70_20_10}
%     %\Description{Whole dataset is combined and split randonly into training, generator training, test}
% \end{figure}
%
%

{\bf Takeaway:} \textit{The attacker can successfully carry out the attack even without access to the training data used for the classifier. For activity recognition models, the attack can be effective with fewer samples; however, more samples are generally needed for a higher success rate in the case of gait recognition. Additionally, we observe that the perturbations are smaller when the attacker uses a larger dataset to train the generator.}
% The Attack success rate does not significantly decrease with a disjoint dataset, indicating that our attack model remains effective even when the threat model includes the additional constraint of the attacker lacking access to the training data of the classifier. This represents a more realistic scenario. Overall, the ASR and MAE trends as a function of trigger data percentage remain consistent.}

% \subsection{Strengthen Attack - Retraining Model}
% Impact of weight parameter to trade off between generator loss and classifier loss

% Does it improve if classifier retraining and without fixing  - Table

% Does discriminator improve stealthiness?

% \subsection{Generaliziblity}
% train generator on the UCI classifier model and test the attack on the Motion-sense dataset (transfer learning / disjoint ??)

% Use the same generator trained using UCI, run Motionsense data (the same number of labels), or GAIT (binary), and see the ASR after running through the destination ASR.

% How to handle different number of features on the generator

% Teacher- student 

% Train on multiple models - CNN/LSTM/NN

% impact on parameters.

\section{Defense Mechanism}
In this section, we assess the robustness of our attack techniques against several state-of-the-art adversarial defense mechanisms. Our goal is to determine how effectively these defenses can mitigate the impact of our trigger-generation approach \cite{zhou2022adversarial}.

% There are ways to defend against adversarial attacks \cite{zhou2022adversarial}. In this section, we evaluate the robustness of our attack techniques against some state-of-the-art adversarial defense mechanisms in suppressing the effectiveness of our trigger-generation technique.
\subsection{Activation clustering}
A common defense mechanism is activation clustering~\cite{chen2018detecting}, which visualizes the activations of the classifier’s hidden layers both with and without the presence of a trigger. By examining these activations, this technique helps identify anomalies or deviations that may indicate the presence of a backdoor trigger. Specifically, it can reveal distinct patterns or clusters in the activations that do not appear in the clean data.
To implement this, we first use t-Distributed Stochastic Neighbor Embedding (t-SNE) to reduce the dimensionality of the activation data. Next, we apply the K-Means clustering technique to form two clusters. The hypothesis is that clean data and backdoor data for a given target label will form separate clusters, indicating the presence of a backdoor attack. 

\begin{figure*}[ht]
    \centering
    \begin{tabular}{cccc}
     \includegraphics[width=1.7in]{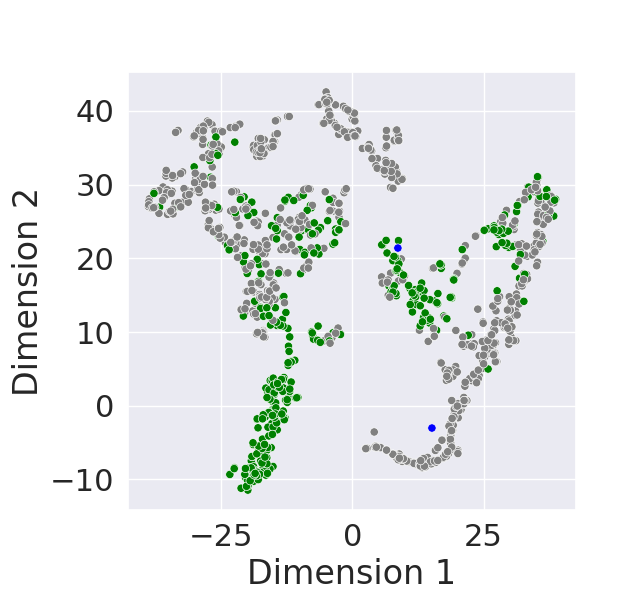}&  \includegraphics[width=1.7in]{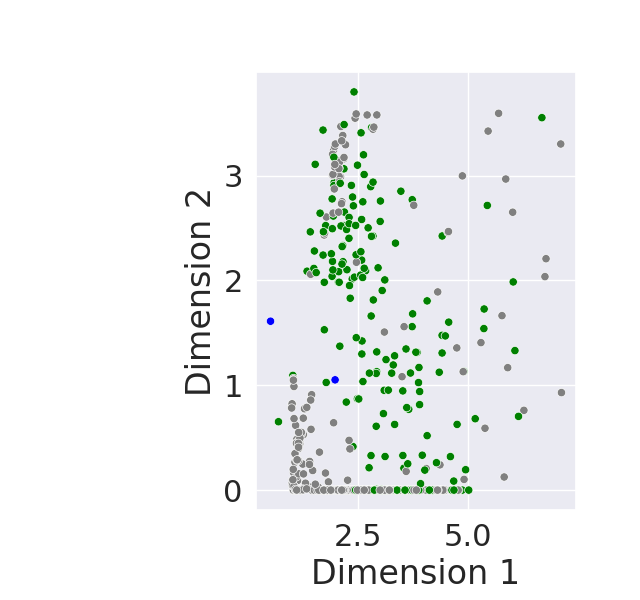}& 
     \includegraphics[width=1.7in]{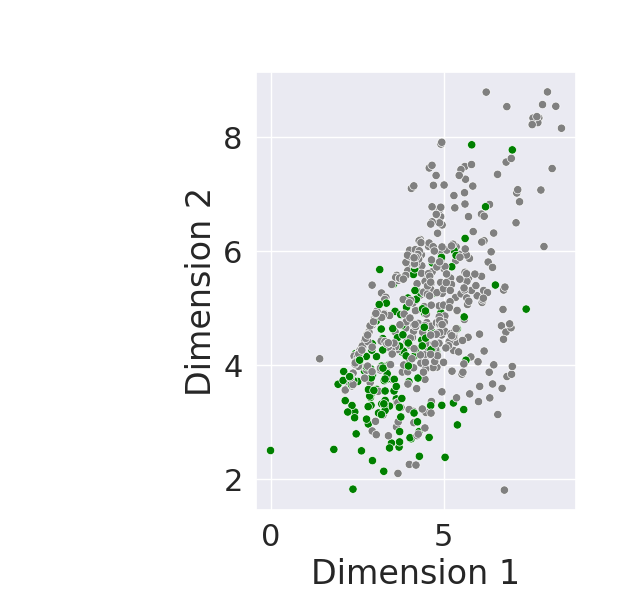}&
     %https://docs.google.com/drawings/d/1zqXfFjzpZcL1mDN-WOCSXDI9XGY-4vXJ4oxAN6k8I9w/edit
     \includegraphics[width=1.2in]{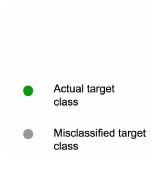}\\ %{figures/Label (3).pdf}\\
     %figures/TSNE GT labels here.pdf}\\
     \footnotesize
     (a) Combined & \footnotesize(b) Activation cluster 1 & \footnotesize(c) Activation cluster 2 &  \\
\end{tabular}
    \caption{Activations of the last hidden layer of the classifier on UCI clean data and attack data with trigger classified as target class (0) are shown in (a). Two 
    activation clusters of target class (0) predictions by the classifier are shown in (b) and (c), indicating no distinct clusters formed by actual target class activity (green) and misclassified target class predictions (grey) by the classifier. Misclassified target class activity predictions due to the presence of trigger are spread across both clusters.} 
    %\caption{t-SNE plot of hidden layer output for benign prediction.} 
    % (a) Combined  (b) Activation Cluster 1 (c) Activation Cluster }
    % 2. Labels are 0: Benign data, 1: Attack data-backdoored, 2: Wiretapping attack data, 3: Mirai attack, 4: Fuzzing attack}
     \label{fig:TSNE}
     %\Description{TSNE}
     \vspace{-3mm}
\end{figure*}

For our evaluation, we cluster all samples, clean and backdoor, with target label 0.
Figure~\ref{fig:TSNE} shows the t-SNE plot with the combined data and highlights the two clusters identified by the K-Means algorithm in the UCI dataset. As shown, the clean and backdoor data do not form distinct clusters, making it challenging to differentiate between clean and backdoor data. This indicates that the activation clustering method fails to effectively identify the presence of a backdoor attack. 

% We use the activation clustering~\cite{chen2018detecting} technique to visualize the activations of the classifier hidden layer with and without the presence of a trigger. When we visualize the T-SNE plot of target class predictions of input data from all classes with or without trigger, we find that activations are not forming clusters corresponding to data with trigger and without trigger, making it indistinguishable. We then use the K-Means clustering technique to form 2 clusters corresponding to target class predictions, and we see both clusters have data belonging to activities with trigger and without triggers. Activation clusters and TSNE plots are shown in Figure~\ref{fig:TSNE}.

\subsection{Pruning}
%https://pytorch.org/tutorials/intermediate/pruning_tutorial.html
%https://openaccess.thecvf.com/content/ICCV2021W/AROW/papers/Jordao_On_the_Effect_of_Pruning_on_Adversarial_Robustness_ICCVW_2021_paper.pdf
%https://arxiv.org/pdf/1803.03635
%
Another defense mechanism is pruning techniques~\cite{jordao2021effect}, which reduce the size of the network by setting the weights of certain layers to zero~\cite{frankle2018lottery}. These techniques are effective because they preserve the accuracy of the classifier on clean data by eliminating irrelevant or redundant structures in the network. We apply pruning methods using the PyTorch library, performing both local and global pruning with 0.5% pruning of the linear layer. 

Table~\ref{tab:Pruning} shows our results, highlighting that while pruning reduces the classifier's accuracy on clean data, it does not enhance the model's resilience to adversarial attacks. Importantly, pruning was performed after training our generator model, demonstrating that our attack remains robust even when the classifier's capacity is reduced.

% Pruning is the process followed in reducing the size of a dense neural network by setting the weights of certain layers to zero~\cite{frankle2018lottery}. Pruning is proven to be an effective defense mechanism against adversarial attacks~\cite{jordao2021effect}.
% The pruning criterion plays a role in preserving accuracy of classifier on clean model after pruning, as it decides which structures are irrelevant or redundant to the network. Depending on the strategy, for measuring importance, the pruning criterion can also become a bottleneck in the computational cost of pruning process.
% We follow two pruning techniques to evaluate the effectiveness of pruning-based defense against adversarial attacks. The impact of local pruning and global pruning with 0.5\% of pruning of linear layer is shown in Table~\ref{tab:Pruning}. Both pruning techniques result in a drop in classifier accuracy on clean data without any gain in terms of defending against adversarial attacks.
% Pruning looks less effective on sparse models.

% \begin{table}[]
% \centering
% \caption{Impact of pruning on our trigger generation technique.}
% \label{tab:Pruning}
% \begin{tabular}{c|c|c|c}
% \toprule
% %\hline
%                    Performance metric & Our Attack & Local pruning & Global pruning \\ \hline
% \midrule
% Attack ASR          & 0.98                           & 0.98          & 0.98           \\ \hline
% Clean Accuracy (\%) & 94.3                           & 92.9          & 93.9           \\ \hline
% %\bottomrule
% \end{tabular}
% \end{table}

\begin{table}[t]
\centering
\caption{Impact of pruning on our trigger generation technique on UCI dataset.}
\label{tab:Pruning}
\begin{tabular}{|c|c|c|c|} 
\hline
Performance metric  & Our Attack & Local pruning & Global pruning  \\ 
\hline
Attack ASR          & 0.98       & 0.98          & 0.98            \\ 
\hline
Clean Accuracy (\%) & 94.3       & 92.9          & 93.9            \\
\hline
\end{tabular}
\end{table}

\subsection{Adversarial training}
We further assess the impact of adversarial training on our trigger-based attack. To do this, we first retrain the pre-trained classifier using samples generated by our trigger model. After adversarial training, we observe an increase in the classifier's accuracy to 94.12\% on adversarial data, as shown in Table~\ref{tab:ad-training}. This suggests that the attack can be mitigated if the defender has access to the attacker's trigger model.

Following this, we freeze the adversarially trained classifier and retrain the generator using the adversarially trained classifier. Despite this, the attack remains successful, with an attack success rate (ASR) of 0.87 for the newly trained generator. These results indicate that vulnerabilities persist in the adversarially trained model, which attackers can still exploit.

\begin{table}[t]
\centering
\caption{Impact of adversarial training on our attack model on Gait dataset.}
\label{tab:ad-training}
\begin{tabular}{|c|c|c|}
\hline
\begin{tabular}[c]{@{}c@{}}\\ \end{tabular} & \begin{tabular}[c]{@{}c@{}}Before adversarial\\ training\end{tabular}
& \begin{tabular}[c]{@{}c@{}}After adversarial\\ training\end{tabular} \\
\hline
Clean Accuracy (\%)  & 95.66  & 94.12 (old generator)  \\
\hline
Attack ASR   & 0.91  & 0.87 (new generator)         \\                                              \hline         
\end{tabular}
\end{table}

%\subsection{Quantization}
%https://dl.acm.org/doi/pdf/10.1145/3453688.3461755
%\subsection {Neural cleanse / Unlearning}

%https://github.com/Trusted-AI/adversarial-robustness-toolbox/blob/main/notebooks/poisoning_defense_neural_cleanse.ipynb

%https://github.com/Trusted-AI/adversarial-robustness-toolbox/blob/main/notebooks/poisoning_defense_activation_clustering.ipynb

%https://arxiv.org/pdf/2405.03918v1

%\subsubsection {Neural cleanse}
%\subsubsection{GAN based adversarial training}
%\subsubsection{Activation clustering}
%Defensive distillation

% \section{Discussions}
% \input{discussion}
\section{Related work}
Sensor-based IoT systems have been extensively studied in the literature, with numerous applications demonstrating their effectiveness~\cite{6365160, zou2020gait}. Recently, deep learning has become increasingly integral to these systems, enhancing their capabilities in tasks like human activity recognition (HAR) and authentication. For instance, deep learning techniques have been applied to sensor-based data, particularly from wearable devices, to monitor daily activities and gain insights into user behavior~\cite{Malekzadeh:2018:PSD:3195258.3195260, malekzadeh2019mobile}. These systems are adept at capturing a wide range of activities, from simple movements to complex behaviors, making them valuable tools for both monitoring and analysis.
Gait-based authentication and feature extraction have also been widely researched, showcasing the potential of using walking patterns as a biometric for security and identification~\cite{zou2020gait}. Our work is relevant to these systems, particularly where deep learning models are employed for various types of analysis. While we focus on three specific datasets, our approach can be applied to a broader range of human activity recognition (HAR) datasets. This includes datasets labeled with daily activities~\cite{anguita2013public, micucci2017unimib} and hand gestures~\cite{chavarriaga2013opportunity, stiefmeier2008wearable}, which remains part of future work.

Adversarial attacks on DNN have been studied extensively in the vision and natural language domain~\cite{goodfellow2014explaining, koh2017understanding, eykholt2018robust, biggio2018wild}. Prior work has proposed sophisticated adversarial attacks based on GAN architecture that can fool the face recognition models using personal accessories such as glasses~\cite{sharif2019general}. 
%Prior work has proposed sophisticated adversarial attacks based on GAN architecture, capable of deceiving face recognition models using personal accessories such as glasses~\cite{sharif2019general}. 
While our work shares similarities in using a GAN-based network, it diverges significantly in its approach. Prior research predominantly assumes a white-box scenario, where the attacker has full access to the training process and model parameters. In contrast, our method operates in a black-box setting, where we do not modify the model parameters or the training process. This makes our approach more challenging and realistic, as it reflects scenarios where the attacker has limited knowledge or access. Additionally, our work is focused on sensor data, which has distinct characteristics compared to vision or natural language data, making the attack vectors and defenses different.

Our technique employs generator-based perturbation to dynamically generate trigger patterns for various inputs in sensor-based recognition models, such as gait-based authentication and human activity recognition. 
While perturbation-based privacy preservation techniques are extensively studied, few works address the offensive security side of it in the IoT sensor data domain~\cite{chathoth2022differentially, malekzadeh2019mobile}.
A closely related work in the vision domain is~\cite{nguyen2020input}, where the authors demonstrate trigger generation on images. However, unlike our approach, their work does not assume a black-box scenario and does not impact the classifier's performance. Additionally, we evaluate the efficacy of various defense techniques proposed in the literature, which have primarily been validated in the vision domain~\cite{carlini2017towards, jordao2021effect, frankle2018lottery}. Our analysis reveals that these defenses are less effective in the context of sensor data, highlighting the need for future research to develop more robust defenses tailored to sensor-based recognition systems.

\section{Conclusion}
In this paper, we demonstrate that deep learning-based sensor IoT systems can be vulnerable to backdoor attacks, even in a black-box setting where the attacker only has access to the predicted target label. To explore this vulnerability, we develop a novel generator-based dynamic trigger generation technique that operates in a black-box scenario. This technique generates perturbations that, when added to the sensor signal, can fool the model into misclassifying the input. We validate the success of our attack on both gait-based authentication and human activity recognition models.
Our results highlight the effectiveness of this approach across various scenarios, emphasizing the need for more robust defenses against such attacks in sensor-based recognition systems. We observe that minimal perturbation to the original signal is sufficient to execute the backdoor attack. These perturbations are dynamic and vary from signal to signal, making it challenging for defenders to identify a consistent trigger pattern. Additionally, we evaluate our method against several standard defense techniques and find that they are largely ineffective in detecting or mitigating our attack.

{\bf Acknowledgement.}
This material is based upon work supported by the Department of Energy under Award Number DE-CR0000041. Neither the United States Government nor any agency thereof, nor any of their employees, makes any warranty, express or implied, or assumes any legal liability or responsibility for the accuracy, completeness, or usefulness of any information, apparatus, product, or process disclosed, or represents that its use would not infringe privately owned rights. The views and opinions of authors expressed herein do not necessarily state or reflect those of the United States Government or any agency thereof.

\bibliographystyle{IEEEtran}
\bibliography{sample-base}
%\end{thebibliography}

% \bibitem{b1} G. Eason, B. Noble, and I. N. Sneddon, ``On certain integrals of Lipschitz-Hankel type involving products of Bessel functions,'' Phil. Trans. Roy. Soc. London, vol. A247, pp. 529--551, April 1955.
% \bibitem{b2} J. Clerk Maxwell, A Treatise on Electricity and Magnetism, 3rd ed., vol. 2. Oxford: Clarendon, 1892, pp.68--73.
% \bibitem{b3} I. S. Jacobs and C. P. Bean, ``Fine particles, thin films and exchange anisotropy,'' in Magnetism, vol. III, G. T. Rado and H. Suhl, Eds. New York: Academic, 1963, pp. 271--350.
% \bibitem{b4} K. Elissa, ``Title of paper if known,'' unpublished.
% \bibitem{b5} R. Nicole, ``Title of paper with only first word capitalized,'' J. Name Stand. Abbrev., in press.
% \bibitem{b6} Y. Yorozu, M. Hirano, K. Oka, and Y. Tagawa, ``Electron spectroscopy studies on magneto-optical media and plastic substrate interface,'' IEEE Transl. J. Magn. Japan, vol. 2, pp. 740--741, August 1987 [Digests 9th Annual Conf. Magnetics Japan, p. 301, 1982].
% \bibitem{b7} M. Young, The Technical Writer's Handbook. Mill Valley, CA: University Science, 1989.

% \vspace{12pt}
% \color{red}
% IEEE conference templates contain guidance text for composing and formatting conference papers. Please ensure that all template text is removed from your conference paper prior to submission to the conference. Failure to remove the template text from your paper may result in your paper not being published.

%\newpage
%\appendix
%\input{results_old}
\end{document}